# Magnetic Order in A Quenched-High-Temperature-Phase of Cu-Doped MnBi

*Gina Angelo,*[a] *Jeremy G. Philbrick,*[b] *Jian Zhang,*[d] *Tai Kong,*[b,c] *Xin Gui* [a*]

[a] Department of Chemistry, University of Pittsburgh, Pittsburgh, PA, 15260, USA
[b] Department of Physics, University of Arizona, Tucson, AZ, 85721, USA
[c] Department of Chemistry and Biochemistry, University of Arizona, Tucson, AZ, 85721, USA
[d] The Molecular Foundry, Lawrence Berkeley National Laboratory, Berkeley, CA, 94720, USA

## *Abstract*

Permanent magnets are of great importance due to their vast applications. MnBi has been proposed to be a potential permanent magnet that can be widely used while past efforts have been focused on optimizing the ferromagnetic low-temperature phase of MnBi. Herein, we report a series of new materials, $Cu_xMn_{1-x}Bi$, crystallizing in a quenched high-temperature-phase (QHTP) MnBi-related structure. We synthesized single crystals of $Cu_xMn_{1-x}Bi$ and found that they crystallize in an unreported trigonal structure (*P* -31*c*). Magnetic properties measurements imply high-temperature antiferromagnetic (AFM) ordering and low-temperature ferromagnetic or ferrimagnetic (FM/FiM) ordering. By analyzing the doping effect on crystal structure and magnetic properties, we established a magnetic phase diagram for Cu-doped MnBi and attributed the AFM and FM/FiM to two different atomic sites of Mn.

## Introduction

Permanent magnets where the ferromagnetic (FM) transition temperature ($T_C$) is above room temperature have been attracting much attention due to their potential applications, such as magnetic sensors[1,2] and spintronic devices[3]. In particular, the rare-earth-free high-temperature ferromagnets are highly demanded due to the availability of rare-earth elements. Various research has been conducted to search for high-performance ferromagnets without rare-earth elements; for example, AlNiCo magnets,[4,5] MnX (X = Al, Ga and Bi)[3,6–8] and FePt, CoPt and FeNi phases[9–11] have been studied. Among them, MnBi is of great interest due to its high $T_C$, large magnetocrystalline anisotropy, positive temperature coefficient of coercivity and large energy product $(BH)_{max}$ which is comparable to the currently widely applied rare-earth-contained permanent magnet, $Nd_2Fe_{14}B$.[12–15]

MnBi undergoes a structural phase transition from a FM low-temperature phase (LTP) crystallizing in a hexagonal space group $P\,6_3/mmc$ to a paramagnetic (PM) quenched high-temperature phase (QHTP) crystallizing in an orthorhombic space group $P\,222_1$ at ~628 K, due to peritectic decomposition.[16,17] Part of the Mn in LTP-MnBi diffuses into an interstitial site and results in the QHTP-MnBi while the FM properties of LTP-MnBi rapidly vanish once the structural phase transition occurs.[16] The existence of the QHTP-MnBi limits the application of MnBi due to the antiferromagnetic (AFM) interaction between the interstitial Mn site and the original Mn site, which weakens the FM coupling in MnBi.[18–20] Many efforts have been made to suppress the peritectic decomposition and optimize the magnetic properties of MnBi, such as Cr-doped MnBi,[21] Ga-doped MnBi[22,23], Fe-doped MnBi[24] and carbon-modified MnBi[25], however, obstacles towards single-phase product still persist. Cu-doped MnBi with LTP-MnBi structure was also investigated in both film and bulk forms.[26,27] However, these studies focused on the LTP-MnBi, i.e., the samples were molten and annealed under low temperature to obtain the desired FM hexagonal LTP-MnBi. To the best of our knowledge, there was no report about enhancing ferromagnetic order for QHTP-MnBi, which can also overcome the current challenges, in another way, for MnBi permanent magnets.

In this paper, we report the synthesis and characterization of a new structure type that is unreported previously for any of the MnBi system. The new structure crystallizing in a trigonal space group is stabilized by Cu doping, i.e., $Cu_xMn_{1-x}Bi$, and is closely related to the reported LTP-MnBi and QHTP-MnBi, which indicates that the trigonal $Cu_xMn_{1-x}Bi$ is a new high-temperature phase (HTP) of MnBi.

Large single crystals of $Cu_xMn_{1-x}Bi$ are synthesized and the magnetic properties are characterized. High-temperature antiferromagnetism was observed along with a ferro- or ferrimagnetic (FiM) region at low temperature. The evolution of magnetic properties was also investigated and correlated with the dopant's concentration. Moreover, a magnetic phase diagram has been proposed for the system. The new structure type of Cu-doped MnBi reported here can help improve the interpretation of structure-property relationship in the MnBi systems.

## *Experimental Details*

**Crystal Growth of $Cu_xMn_{1-x}Bi$:** Single crystals of $Cu_xMn_{1-x}Bi$ (x = 3%, 4%, 8%, 11% and 14%) were synthesized via Bi flux where Cu powder (99.5%, -150 mesh, Alfa Aesar), Mn powder (99.3%, -325 mesh, Alfa Aesar), and Bi powder (99.5%, -100 mesh, Thermo Scientific) were placed in an alumina crucible and sealed in an evacuated quartz tube. The ratio of Cu, Mn, and Bi were (0.3, 0.5, 0.8, 1.2 1.5):7:15. The tubes were heated at a rate of 180˚C/hr to 1000˚C for 48hr, then cooled at a rate of 3˚C/hr to 400˚C and were subsequently centrifuged. Shiny metallic needle-like crystals of Cu doped MnBi were obtained, as shown in Figure 1(d).

**Phase Identification:** Crystals of $Cu_xMn_{1-x}Bi$ were powdered and analyzed for phase identification via powder X-ray diffraction (PXRD). A Bruker D2 PHASER was used with Cu Kα radiation (λ = 1.54060 Å, Ge monochromator). The Bragg angle measured was from 5 to 90° at a rate of 0.6°/min with a step of 0.01°. Rietveld fitting in FULLPROF was employed to test the phase purity of $Cu_xMn_{x-1}Bi$.[28]

**Crystal Structure Determination:** Single crystal samples of $Cu_xMn_{1-x}Bi$ were cut into a suitable size and mounted onto MiTeGen MicroMounts™ for low temperature (150 K) measurement. Single crystal X-ray diffraction data was collected using Bruker D8 VENTURE equipped with the IμS 3.0 Mo Kα radiation. Indexing was performed using APEX4 (Difference Vectors method). Space groups were determined using XPREP implemented in APEX4. A numerical absorption correction based on crystal-face-indexing was applied using *XPREP*. The direct method and full-matrix least-squares on $F^2$ procedure within the SHELXTL package were employed to solve the crystal structure. [29,30] Diffraction patterns from single crystal XRD are shown in Figures S1-S5 in the Supporting Information.

**Magnetic Properties Measurements:** Magnetization was measured in a Quantum Design physical property measurement system (PPMS) dynacool (1.8- 300 K, 0- 90 kOe) equipped with a vibrating sample magnetometer. Single crystals were mounted on a silica sample holder in preferred orientations using GE varnish.

## *Results and Discussion*

**Crystal Structure of $Cu_xMn_{1-x}Bi$ and Structural Relationship between MnBi and $Cu_xMn_{1-x}Bi$:** Crystals of $Cu_xMn_{1-x}Bi$ (x = 3%, 4%, 8%, 11% and 14%) are synthesized and characterized. Crystals did not form when the ratio of Cu was less than 0.3, relative to Cu:Mn:Bi = 0.3:7:15. All five $Cu_xMn_{1-x}Bi$ samples were determined to crystallize in the same structure with centrosymmetric trigonal space group *P* -31*c* (No. 163), as shown in Figure 1(a), by using single crystal X-ray diffraction (SCXRD). The refined crystallographic data collected under 150 K with atomic positions, site occupancies and refined anisotropic displacement parameters are listed in Tables 1 and Table S1 & S2 in Supporting Information. The chemical compositions of $Cu_xMn_{1-x}Bi$ were identified as $Cu_{0.03(2)}Mn_{0.97(2)}Bi_{0.99(1)}$, $Cu_{0.04(2)}Mn_{0.96(2)}Bi_{0.99(1)}$, $Cu_{0.08(2)}Mn_{0.92(2)}Bi_{0.99(1)}$, $Cu_{0.11(4)}Mn_{0.89(4)}Bi_{0.99(1)}$ and $Cu_{0.14(7)}Mn_{0.86(7)}Bi_{0.99(1)}$. Two crystallographically equivalent sites are found for both Mn and Bi atoms while the dopant, Cu, locates at Mn sites. Depending on the concentration of Cu, it can be observed on different Mn sites. As can be seen in Figure 1(a), single-site doping (SSD) of Cu occurs when $x \lesssim 8\%$ and Cu mixes with Mn2 site (Wyckoff position: 2*d*). Double-site doping (DSD) takes place when $x \gtrsim 11\%$ while partial occupancy of Cu can be found on both Mn1 (6*g*) and Mn2 (2*d*) sites. Meanwhile, site splitting was found for Bi2 and Mn2 site in all samples. The crystal structure of $Cu_xMn_{1-x}Bi$ is unreported and distinct from both previously reported LTP- and QHTP- MnBi, i.e., it is a new structure. The Mn framework of $Cu_xMn_{1-x}Bi$ is shown in Figure 1(b). A hexagonal channel framework of Mn(Cu) exists when viewing along *c* axis while a three-dimensional (3-D) Kagome lattice of Mn(Cu) can be found from the side view. Two different types of coordination are identified in $Cu_xMn_{1-x}Bi$, as shown in Figure 1(c), which include Mn1@$Bi_6$ tilted octahedron and Mn2@$Bi_5$ trigonal bipyramid.

The structural relationship between previously reported MnBi and $Cu_xMn_{1-x}Bi$ is shown in Figure 2. As mentioned before, the permanent magnet MnBi (S.G.: *P* $6_3$/*mmc*, No. 194, LTP) undergoes a structural phase transition at ~628 K upon heating where Mn diffuses to the interstitial 2*d* site and forms another structure (*P* $222_1$, No. 17, QHTP).[16] The Mn atoms construct a triangular lattice stacking along the *c* axis in MnBi-LTP while the symmetry of the unit cell changes in MnBi-QHTP due to the occurrence of the interstitial Mn. However, in $Cu_xMn_{1-x}Bi$, part of the Mn atoms in MnBi-LTP triangular lattice vanish and thus, Mn1 site (Wyckoff position: 6*g*) of Cu-doped MnBi assembles a Kagome lattice stacking along the *c* axis while the Mn2 atoms occupy the 2*d* site. The missing Mn on 2*b* site (atomic site (0,0,0) in $Cu_xMn_{1-x}Bi$) and the 2*d* Mn2 site leads to the disappearance of six-fold symmetry of Mn lattice but results in a three-fold symmetry. Interestingly, if QHTP-MnBi can be seen as an outcome of partial Cu diffusion from the 2*a* site of LTP-MnBi, then $Cu_xMn_{1-x}Bi$ can be treated as a result of partial Cu diffusion onto the 2*d* site. Thus, $Cu_xMn_{1-x}Bi$ can be regarded as a new high-temperature phase of MnBi. A comparison of powder X-ray diffraction (PXRD) patterns between $Cu_xMn_{1-x}Bi$ and QHTP-MnBi is shown in Figure S6. Significant differences can be observed in the patterns, which indicates that $Cu_xMn_{1-x}Bi$ adopts different crystal structures.

**Phase Purity and the Effect of Double-Site Doping (DSD):** PXRD patterns collected from 5° to 90° for crushed $Cu_xMn_{1-x}Bi$ crystals are listed in Figure 3(a). Rietveld refinement was conducted for all patterns with the inclusion of Bi as a secondary phase. The calculated weight percentage of Bi ranges from 5 wt.% to 10 wt.% as the Bi flux is unavoidable when making PXRD samples. The $\chi^2$ ($R_p/R_{wp}$) factor for refined $Cu_xMn_{1-x}Bi$ (x = 3%, 4%, 8%, 11% and 14%) are 2.46 (4.69 %/5.97 %), 3.01 (4.37 %/5.86 %), 3.78 (5.36 %/7.21 %), 3.55 (5.43 %/7.08 %) and 3.95 (4.79 %/6.30 %), which indicate reasonable and reliable fitting between the observed and calculated patterns. Figure 3(b) shows the changes of three major Bragg peaks of $Cu_xMn_{1-x}Bi$, i.e., (201), (002) and (210) peaks. The (201) and (210) peaks move to higher 2θ when x ≲ 8% while a dramatic decrease in 2θ can be observed with higher x, i.e., x ≳ 11%. The (002) peaks show different behavior since they shift monotonically to higher 2θ with increasing Cu concentration.

To investigate the non-monotonic behavior of the Bragg peak shift, the lattice parameters of $Cu_xMn_{1-x}Bi$ obtained from single-crystal and powder XRD refinements are summarized in Figure 4(a). For x ≲ 8%, *a* slightly decreases from 8.6584 Å to 8.6447 Å (PXRD; 300 K) and from 8.6159 Å to 8.6059 Å (SCXRD; 150 K) while *c* also shrinks from 5.8432 Å to 5.8341 Å (PXRD; 300 K) and from 5.8162 Å to 5.8059 Å (SCXRD; 150 K). However, dramatic changes can be seen when x ≳ 11% where *a* increases to 8.6962 Å (PXRD; 300 K) and 8.6617 Å (SCXRD; 150 K); *c* drops to 5.7705 Å (PXRD; 300 K) and 5.7386 Å (SCXRD; 150 K). Therefore, the aforementioned non-monotonic behavior of PXRD patterns originates from the change of the trend of lattice parameters due to the second site doping of Cu on Mn 2*d* site when x ≳ 11%.

The relationship between x in $Cu_xMn_{1-x}Bi$ and *c*/*a* ratio is demonstrated in Figure 4(b). No significant difference of *c*/*a* can be seen under 150 K and 300 K while it drops for ~0.6% and ~1.2 % from ~0.675 to ~0.671 and to ~0.663 after DSD is observed at x $\gtrsim$ 11%. Interestingly, V becomes larger when x = 15%, although the unit cell is expected to shrink when larger Mn (atomic radius ~ 1.40 Å) is substituted by smaller Cu (~ 1.35 Å).[31] The influences of Cu-doping on Mn-Mn interatomic distances are also illustrated in the top of Figure 4(c). All the bond lengths are collected from SCXRD data measured under 150 K and the corresponding Mn1-Mn1 and Mn1-Mn2 bonds are shown in Figure 4(d). When x $\lesssim$ 8%, $d_{Mn1\text{-}Mn1}$, i.e., the distance between undoped Mn1 sites, decreases insignificantly from 2.9081 Å (x = 3%) to 2.9030 Å (x = 8%). However, when DSD occurs, $d_{Mn1\text{-}Mn1}$ drops to 2.8937 Å (x = 11%) and eventually to 2.8693 Å (x = 14%). On the other hand, $d_{Mn1\text{-}Mn2}$, i.e., the distance between two nearest crystallographically inequivalent Mn sites, decreases from 2.8810 Å (x = 3%) to 2.8773 Å (x = 8%). Unlike $d_{Mn1\text{-}Mn1}$, $d_{Mn1\text{-}Mn2}$ then increases to 2.8779 Å (x = 11%) and 2.8828 Å (x = 14%). Such DSD induced anomalous changes on Mn-Mn distances can have significant impact on the magnetic interaction between Mn atoms, which will be discussed later.

**High-Temperature Antiferromagnetism in $Cu_xMn_{1-x}Bi$:** Magnetic properties of $Cu_xMn_{1-x}Bi$ were measured under two different directions of external magnetic field (H), i.e., H is parallel to *c* axis of the sample (H//*c*) and H is perpendicular to the *c* axis (H⊥*c*). Figures 5(a) & 5(c) present the temperature-dependence of magnetic susceptibility ($\chi$) measured under $\mu_0H$ = 0.1 T for both magnetic field directions and the inverse $\chi$ vs T curves are shown in Figures 5(b) & 5(d). Magnetic anisotropy was observed due to the significant difference in $\chi$ for both field directions. From Figure 5(a), when H//*c*, antiferromagnetic (AFM) transition peaks can be seen within the range of 200 K-260 K for all five samples. The AFM peaks appear to be at $T_{N1}$ ~235 K (x = 3%), ~242 K (x = 4%), ~253 K (x = 8%), ~243 K (x = 11%) and ~209 K (x = 14%). Within SSD range, AFM transitions take place at higher temperatures with larger x while the peaks shift to lower temperatures when DSD occurs. The AFM peak positions observed in Figure 5(a) are marked in Figure 5(b) and labeled as AFM1 peaks. Similarly, AFM peaks also emerge while cooling when H⊥*c*. The major AFM transition peaks happen at lower temperatures compared to H//*c*, i.e., $T_{N2}$ ~211 K (x = 3%), ~217 K (x = 4%), ~227 K (x = 8%), ~215 K (x = 11%) and ~173 K (x = 14%). The peaks corresponding to $T_{N2}$ are labeled as AFM2 peaks and marked in Figure 5(d). Since AFM1 and AFM2 peaks exist at different temperatures, they are expected to be two different magnetic features for all samples because in both Figures 5(a) and 5(c), both a kink and a peak can be observed at both $T_{N1}$ and $T_{N2}$. Therefore, the AFM1 peaks should be a PM to AFM transition while AFM2 peaks are likely to correspond to a spin reorientation or a different AFM ordering transition for another Mn sublattice. A temperature-dependent heat capacity measurement has been carried out for a representative sample, $Cu_{0.04}Mn_{0.96}Bi_{0.99}$, from 180 K to 270 K, as shown in Figure S7. Two transitions can be observed near ~ 241 K and ~217 K, which are consistent with $T_{N1}$ and $T_{N2}$ determined by magnetic measurements.

While $Cu_xMn_{1-x}Bi$ are further cooled down, clear FM-like transitions take place when x $\lesssim$ 8%, as can be seen in Figures 5(a) & 5(c) where the transition temperatures ($T_C$) are consistent in both directions of magnetic field. With increasing x, $T_C$ decreases from ~149 K (x = 3%) to ~114 K (x = 4%) and ~74 K (x = 8%). A broad FM-like transition peak can also be observed at ~34 K for x = 11% and ~45 K for x = 14%, determined from $d(\chi T)/dT$ *vs* T curves shown in Figure S8. To identify the magnetic ordering type below $T_C$, field-dependence of magnetization of $Cu_xMn_{1-x}Bi$ is required and it will be discussed in the next section.

Curie-Weiss (CW) fitting is performed to obtain information about the paramagnetic state of $Cu_xMn_{1-x}Bi$. The following CW law was applied to fit all the curves from 275 K to 300 K:

$$\chi = \frac{C}{T-\theta_{CW}},$$

where $\chi$ is the magnetic susceptibility of the material, C is a constant independent of temperature and related to the effective moment ($\mu_{eff}$) and $\theta_{CW}$ is the Curie-Weiss constant. The fitting results are shown in Figure 6. The trends for $\mu_{eff}$ and $\theta_{CW}$ are consistent for both H//*c* and H$\perp$*c* where $\theta_{CW}$ ($\mu_{eff}$) appears to monotonically decrease (increase) with more Cu. The fitted positive $\theta_{CW}$ for all samples indicate the dominant ferromagnetic interaction within the paramagnetic region at high temperature. For $\mu_{eff}$, a transition from $Mn^{3+}$ to $Mn^{2+}$ with evolving Cu concentration is seen based on the spin-only moment for $Mn^{3+}$ (~4.90 $\mu_B$) and $Mn^{2+}$ (~5.92 $\mu_B$) based on $\mu_{eff}$ (spin-only) = $\sqrt{n(n+2)}$. Such a change of the valence state of Mn is anomalous. Considering that Mn starts with $Mn^{3+}$ in $Cu_{0.03}Mn_{0.97}Bi_{0.99}$, it is expected to possess a higher oxidation state, such as $Mn^{4+}$, to compensate for the inclusion of Cu, which is usually considered to be divalent ($d^9$ configuration with $\mu_{eff}$ = 1.73 $\mu_B$). Thus, $\mu_{eff}$ is expected to decrease with increase of x, which is inconsistent with what is observed for $Cu_xMn_{1-x}Bi$. Therefore, further investigation will be needed to fully understand this anomalous behavior. Moreover, a correlation between $\theta_{CW}$ and $d_{Mn1-Mn1}$ can be observed by comparing Figure 6 with Figure 4(c). While $d_{Mn1-Mn1}$ decreases monotonically with higher concentration of Cu, $\theta_{CW}$ also drops with more Cu. It implies that Mn1-Mn1 interatomic distances might be the key factor for FM interaction in $Cu_xMn_{1-x}Bi$.

**Low-Temperature Ferromagnetic Components and Metamagnetic Transitions in $Cu_xMn_{1-x}Bi$:** To further identify the FM-like transitions appeared below ~150 K, magnetic hysteresis loops were measured for $Cu_xMn_{1-x}Bi$ with two directions of external magnetic field, as shown in Figure 7. Strong magnetic anisotropy is observed as the hysteresis loops exhibit differently between H//*c* and H⊥*c*. According to Figure 7(a), when H//*c* and T = 2 K, hard FM behavior can be seen in all samples. The coercive fields ($H_C$) are ~1.1 T (x = 3%), ~1.0 T (x = 4%), ~0.6 T (x = 8%), ~0.9 T (x = 11%) and ~1.6 T (x = 14%) while the remanent magnetizations ($M_r$) are ~1.5 $\mu_B$/Mn (x = 3%), ~1.4 $\mu_B$/Mn (x = 4%), ~1.2 $\mu_B$/Mn (x = 8%), ~1.1 $\mu_B$/Mn (x = 11%) and ~1.3 $\mu_B$/Mn (x = 14%). The magnetic moments do not seem to saturate at 9 T for all samples while the magnetizations at 9 T ($M_{9T}$) are far below the expected values for both $Mn^{2+}$ and $Mn^{3+}$. Therefore, the ground state of $Cu_xMn_{1-x}Bi$ can be either FM, where magnetic saturation is not achieved yet at $\mu_0H$ = 9 T, or FiM where the magnetic moments are cancelled out due to the antiparallelly aligned spins which causes the small $M_{9T}$. The coercivity observed at 2 K indicates predominant FM components in all materials. Based on the field-dependence of magnetization curves at 2 K shown in Figure 8(a), at low magnetic field, small slopes can be seen for all materials. It indicates that the FM domain growth is slow at low field, which is typical for hard ferromagnets. At a higher field, the magnetization appears to grow rapidly for $Cu_xMn_{1-x}Bi$ samples, which implies the rotation of magnetic domains.

When the temperature rises to 100 K, as can be seen in Figure 7(a), FM-like behavior can also be observed but the coercivity is dramatically weakened for $Cu_{0.03}Mn_{0.97}Bi_{0.99}$ and $Cu_{0.04}Mn_{0.96}Bi_{0.99}$. However, for x ≥ 8%, metamagnetic transitions can be found at high magnetic fields, which indicate antiferromagnetically dominant ordering at this temperature region for heavily doped samples. The insets in Figure 7(a) present the weak coercivity observed for x ≥ 8% samples, which might originate from canted AFM order. Similar behavior can be seen for T = 195 K for x = 8% and x = 11%, indicating the same canted AFM state. However, for x = 14% at 195 K, the metamagnetic transition and coercivity become much broader than x = 8% and x = 11%. This might be because the AFM state corresponding to AFM2 peak (174 K for x = 14%) shown in Figure 5(d) disappears due to the elevated temperature and transitions to the AFM state corresponding to AFM1 peaks. Thus, distinct field-dependence of magnetization different than x = 8% and 11% samples is shown. Such broader metamagnetic transitions can be found in all samples except for x = 14 % at T = 220 K, implying the same AFM1 state as x = 14% at 195 K. Based on the insets of Figure 7(a), weak coercivity can also be observed for AFM1 phase, which indicates that AFM1 is slightly canted as well. At room temperature (T = 300 K), all $Cu_xMn_{1-x}Bi$ materials show linear behavior in field-dependent magnetization from 0 to 9 T, which corresponds to PM state.

When H⊥*c*, comparing smaller $M_{9T}$ observed for all samples suggest that the *c* axis is the easy axis of $Cu_xMn_{1-x}Bi$. Obvious FM components, i.e., large slope at low magnetic field or coercivity, can be seen only for all samples at 2 K and $Cu_{0.03}Mn_{0.97}Bi_{0.99}$ and $Cu_{0.04}Mn_{0.96}Bi_{0.99}$ at 100 K Thus, the FM-like transition peaks shown in Figure 5 are related to the low-temperature FM components while such FM components disappear for x ≥ 8% samples because the FM-like transitions for these compounds occur below 100 K.

The metamagnetic transitions at T = 2 K in $Cu_xMn_{1-x}Bi$ are illustrated in Figure 8 by the field-dependent magnetization curves (Figures 8(a) and 8(c)) and their first derivatives (Figures 8(b) and 8(d)). When H//*c*, no major metamagnetic transitions can be observed for x = 8%, 11% and 14%. The only peaks shown in Figure 8(b) for x = 8%, 11% and 14%stand for the rotation of FM domains, as discussed before. However, one metamagnetic transition peak can be found after the emergence of rotation of FM domains for $Cu_{0.03}Mn_{0.97}Bi_{0.99}$ and two metamagnetic peaks are observed before and after the occurrence of the domain wall rotation, respectively. When H⊥*c*, the broad peaks shown in Figure 8(d) for x ≥ 8% samples indicate correspond with the FM domain wall rotation. However, such broad peaks disappear for lightly doped samples. Instead, one or two metamagnetic transition peaks can be observed for x = 3% and 4% materials, respectively.

**Magnetic Phase Diagram of $Cu_xMn_{1-x}Bi$:** To better understand the complicated magnetism in $Cu_xMn_{1-x}Bi$, a magnetic phase diagram is needed and shown in Figure 9(a). Based on the previous discussion, we conclude that there are four magnetic states existing in this system below 300 K, which are paramagnetic (PM), the first and the second canted antiferromagnetic phase (canted AFM1 and canted AFM2), and the low-temperature ferromagnetic/ferrimagnetic state (FM/FiM). Therefore, three sets of data points for corresponding magnetic phases are extracted from Figure S8 and Figure 5. The estimated phase diagram for $Cu_xMn_{1-x}Bi$ system is shown in Figure 9(a). The boundaries between different states are obtained by fitting each dataset by using the parabolic function $y = A + Bx + Cx^2$ where A, B and C are constants. The function is solely used as guide for eyes.

As the phase diagram is constructed, a brief discussion about the relationship between magnetic states and Cu doping shall be made here. Based on the phase diagram, the two sets of AFM transition temperatures possess opposite trends than the FM/FiM one. For example, $T_C$ drops when $T_{N1}$ and $T_{N2}$ increase and $T_C$ increases when $T_{N1}$ and $T_{N2}$ decrease. According to Figure 4(c) where $d_{Mn1\text{-}Mn2}$ decreases from x = 3% to x = 8% and then becomes longer when x = 11% and 14%, it can be found that the low-temperature FM component has analogous trend with $T_C$. Therefore, it is possible that the FM interaction predominantly originates from the interaction between Mn1 and Mn2 sublattices. Meanwhile, as shown in Figure 9(b), it can also be found that the ratio of $T_{N1}$ and $T_{N2}$, i.e., $T_{N1}/T_{N2}$, remains similar when SSD dominates while it increases after DSD occurs. Considering that the occupancy of the Mn2 site exhibits similar trend as the AFM1/AFM2 peaks, we can speculate that Mn2 sublattice contributes to both AFM1 and AFM2 peaks. When DSD takes place, the two AFM peaks diverge due to the change of the exchange interaction pathway, i.e., Mn2-Mn1-Mn2. At low temperature, Mn1 sublattice becomes FM/FiM ordered. With more Cu doped on Mn2 site, the FM/FiM coupling between neighboring Mn1 chains becomes weaker where the non-magnetic impurity, Cu, is the most significant factor. However, when the concentration of Cu continues to increase, $d_{Mn1\text{-}Mn1}$ drops dramatically and thus, the FM/FiM interaction gets stronger, i.e., $T_C$ increases for x = 14%. Here, the interatomic distance between Mn1 atoms becomes the key factor to determine the magnetic interaction type of Mn1-Mn1.

## *Conclusions*

To summarize, a series of new high-temperature phase Cu-doped MnBi was synthesized and characterized. The new structure is found to be distinct from the reported LTP-MnBi and QHTP-MnBi. Two crystallographically inequivalent Mn sites were identified, and Cu doping was observed to occur *via* two different ways: single-site doping (SSD) and double-site doping (DSD). We have noticed significant changes in lattice parameters and Mn-Mn bonds after DSD takes place. Magnetic properties characterizations indicate strong magnetic anisotropy. Meanwhile, $Cu_xMn_{1-x}Bi$ undergoes two antiferromagnetic transitions at higher temperatures while a ferromagnetic-like transition can be observed at lower temperatures. A magnetic phase diagram was also generated. By comparing with the magnetic properties and the doping effect on the crystal structure, it was speculated that AFM states originate from the Mn2 site while the FM interaction is mainly from the Mn1 sublattice.

## *Author Information*

Corresponding Author: xig75@pitt.edu

Notes: The authors declare no competing financial interest.

## *Acknowledgements*

G.A. and X.G. are supported by the startup fund from the University of Pittsburgh and the Pitt Momentum Fund. Work done at the University of Arizona was supported by the University of Arizona startup fund. Work at the Molecular Foundry was supported by the Office of Science, Office of Basic Energy Sciences, of the U.S. Department of Energy under Contract No. DE-AC02-05CH11231.

## References

1. H. Guo, F. Ju, Y. Cao, F. Qi, D. Bai, Y. Wang and B. Chen, *Sens. Actuators, A*, 2019, **285**, 519–530.
2. H. G. Kortier, J. Antonsson, H. M. Schepers, F. Gustafsson and P. H. Veltink, *IEEE Trans. Neural Syst. Rehabil. Eng.*, 2015, **23**, 796–806.
3. L. Zhu and J. Zhao, *Appl. Phys. A*, 2013, **111**, 379–387.
4. S. S. Maroufian and P. Pillay, *IEEE Trans. Ind. Appl.*, 2019, **55**, 4733–4742.
5. L. Zhou, M. K. Miller, P. Lu, L. Ke, R. Skomski, H. Dillon, Q. Xing, A. Palasyuk, M. R. McCartney, D. J. Smith, S. Constantinides, R. W. McCallum, I. E. Anderson, V. Antropov and M. J. Kramer, *Acta Mater.*, 2014, **74**, 224–233.
6. J. H. Park, Y. K. Hong, S. Bae, J. J. Lee, J. Jalli, G. S. Abo, N. Neveu, S. G. Kim, C. J. Choi and J. G. Lee, *J. Appl. Phys.*, 2010, **107**, 09A731.
7. L. Pareti, F. Bolzoni, F. Leccabue and A. E. Ermakov, *J. Appl. Phys.*, 1986, **59**, 3824–3828.
8. Y.-C. Chen, G. Gregori, A. Leineweber, F. Qu, C.-C. Chen, T. Tietze, H. Kronmüller, G. Schütz and E. Goering, *Scr. Mater.*, 2015, **107**, 131–135.
9. J. Zhou, R. Skomski, X. Li, W. Tang, G. C. Hadjipanayis and D. J. Sellmyer, *IEEE Trans. Magn.*, 2002, **38**, 2802–2804.
10. O. D. Oniku, B. Qi and D. P. Arnold, *J. Appl. Phys.*, 2014, **115**, 17E521.
11. A. M. Montes-Arango, L. G. Marshall, A. D. Fortes, N. C. Bordeaux, S. Langridge, K. Barmak and L. H. Lewis, *Acta Mater.*, 2016, **116**, 263–269.
12. G. Bai, R. W. Gao, Y. Sun, G. B. Han and B. Wang, *J. Magn. Magn. Mater.*, 2007, **308**, 20–23.
13. S. N. Piramanayagam, M. Matsumoto and A. Morisako, *J. Magn. Magn. Mater.*, 2000, **212**, 12–16.
14. D. Brown, B.-M. Ma and Z. Chen, *J. Magn. Magn. Mater.*, 2002, **248**, 432–440.
15. J. Park, Y.-K. Hong, J. Lee, W. Lee, S.-G. Kim and C.-J. Choi, *Metals*, 2014, **4**, 455–464.
16. T. Chen and W. Stutius, *IEEE Trans. Magn.*, 1974, **10**, 581–586.

17. J. Mohapatra and J. P. Liu, in *Handbook of Magnetic Materials*, ed. E. Brück, Elsevier, Amsterdam, Netherlands, 2018, vol. 27, pp. 1–57.
18. T. J. Williams, A. E. Taylor, A. D. Christianson, S. E. Hahn, R. S. Fishman, D. S. Parker, M. A. McGuire, B. C. Sales and M. D. Lumsden, *Appl. Phys. Lett.*, 2016, **108**, 192403.
19. Taylor, A. E.; Berlijn, T.; Hahn, S. E.; May, A. F.; Williams, T. J.; Poudel, L.; Calder, S.; Fishman, R. S.; Stone, M. B.; Aczel, A. A.; Cao, H. B.; Lumsden, M. D.; Christianson, A. D. Influence of Interstitial Mn on Magnetism in the Room-Temperature Ferromagnet $Mn_{1+\delta}Sb$. *Phys. Rev. B* **2015**, *91* (22), 224418. https://doi.org/10.1103/PhysRevB.91.224418.
20. T. R. Paudel, B. Lama and P. Kharel, *RSC Adv*, 2021, **11**, 30955–30960.
21. K. Anand, N. Christopher and N. Singh, *Appl. Phys. A*, 2019, **125**, 870.
22. Y. Yang, J.-W. Kim, P.-Z. Si, H.-D. Qian, Y. Shin, X. Wang, J. Park, O. L. Li, Q. Wu, H. Ge and C.-J. Choi, *J. Alloys Compd.*, 2018, **769**, 813–816.
23. C. Zhou and M. Pan, *Int. J. Electrochem. Sci.*, 2019, **14**, 10387–10395.
24. Y. Yang, J. T. Lim, H.-D. Qian, J. Park, J.-W. Kim, O. L. Li and C.-J. Choi, *J. Alloys Compd.*, 2021, **855**, 157312.
25. R. Hirian, R. Dudric, O. Isnard, K. Kuepper, M. Coldea, L. Barbu-Tudoran, V. Pop and D. Benea, *J. Magn. Magn. Mater.*, 2021, **532**, 167997.
26. V. V. Ramakrishna, S. Kavita, R. Gautam, T. Ramesh and R. Gopalan, *J. Magn. Magn. Mater.*, 2018, **458**, 23–29.
27. H. Göbel, E. Wolfgang and H. Harms, *Phys. Status Solidi A*, 1976, **34**, 553–564.
28. J. Rodríguez-Carvajal, *Physica B*, 1993, **192**, 55–69.
29. G. M. Sheldrick, *Acta Cryst.*, 2015, **71**, 3–8.
30. N. Walker and D. Stuart, *Acta Cryst.*, 1983, **39**, 158–166.
31. J. C. Slater, *J. Chem. Phys.*, 2004, **41**, 3199–3204.

**Table 1.** Single crystal structure refinement for $Cu_xMn_{1-x}Bi$ at 150 (2) K.

| Refined Formula | $Cu_{0.03(2)}Mn_{0.97(2)}Bi_{0.99}$ | $Cu_{0.04(2)}Mn_{0.96(2)}Bi_{0.99}$ | $Cu_{0.08(2)}Mn_{0.92(2)}Bi_{0.99}$ | $Cu_{0.11(5)}Mn_{0.89(5)}Bi_{0.99}$ | $Cu_{0.14(5)}Mn_{0.86(5)}Bi_{0.99}$ |
|---|---|---|---|---|---|
| F.W. (g/mol) | 261.52 | 261.12 | 261.72 | 261.99 | 262.49 |
| Space group; Z | *P* -31*c*; 8 | *P* -31*c*; 8 | *P* -31*c*; 8 | *P* -31*c*; 8 | *P* -31*c*; 8 |
| *a* (Å) | 8.6159 (2) | 8.6110 (1) | 8.6059 (2) | 8.6177 (2) | 8.6617 (2) |
| *c* (Å) | 5.8162 (2) | 5.8073 (1) | 5.8059 (2) | 5.7874 (3) | 5.7386 (2) |
| V ($Å^3$) | 373.91 (2) | 372.92 (1) | 372.39 (2) | 372.22 (2) | 372.86 (2) |
| Extinction Coefficient | 0.0022 (3) | 0.0018 (3) | 0.0052 (6) | 0.0012 (2) | 0.0031 (4) |
| θ range (°) | 2.730-33.147 | 2.731-30.531 | 2.733-30.551 | 2.729-33.161 | 2.715-29.000 |
| No. reflections; $R_{int}$ | 7847; 0.0566 | 6321; 0.0518 | 6046; 0.0666 | 8221; 0.0460 | 5185; 0.0718 |
| No. independent reflections | 475 | 389 | 388 | 476 | 339 |
| No. parameters | 22 | 22 | 22 | 23 | 23 |
| $R_1$: $\omega R_2$ ($I$>2δ($I$)) | 0.0220; 0.0530 | 0.0223; 0.0576 | 0.0246; 0.0615 | 0.0210; 0.0492 | 0.0311; 0.0734 |
| Goodness of fit | 1.164 | 1.407 | 1.220 | 1.198 | 1.127 |
| Diffraction peak and hole ($e^-$/ $Å^3$) | 1.743; -1.937 | 1.606; -2.271 | 1.866; -2.289 | 2.008; -1.923 | 1.915; -2.312 |

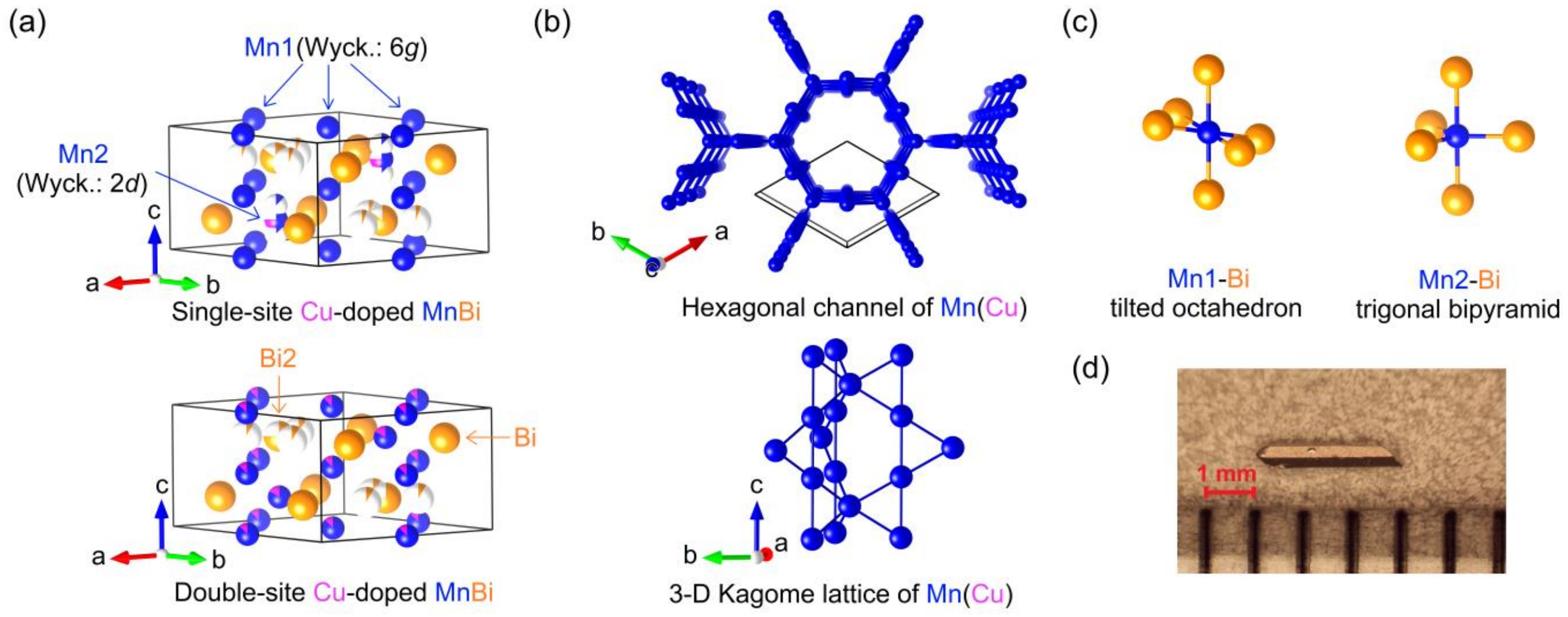

**Figure 1. (a)** Crystal structure of single-site Cu-doped and double-site Cu-doped MnBi. Magenta, blue and orange spheres stand for Cu, Mn and Bi atoms, respectively. **(b)** Mn framework in Cu-doped MnBi. **(c)** Two types of Mn-Bi coordination in Cu-doped MnBi. **(d)** Picture of a representative needle-like Cu-doped MnBi crystal.

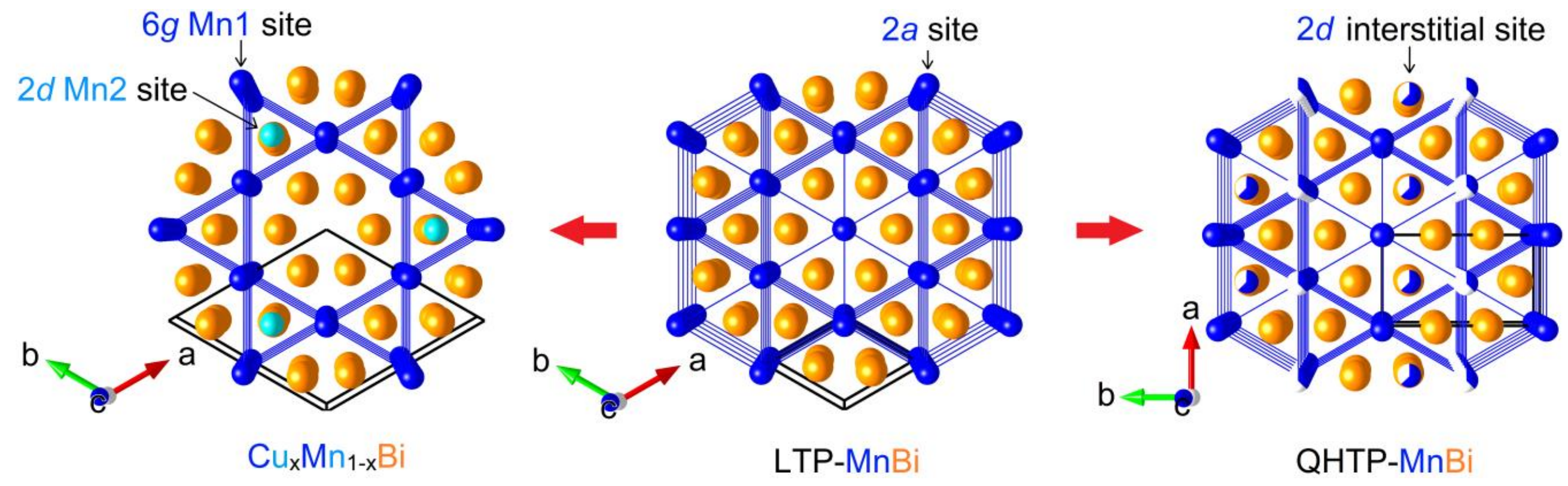

**Figure 2.** Structural relationship between $Cu_xMn_{1-x}Bi$, LTP-MnBi and QHTP-MnBi.

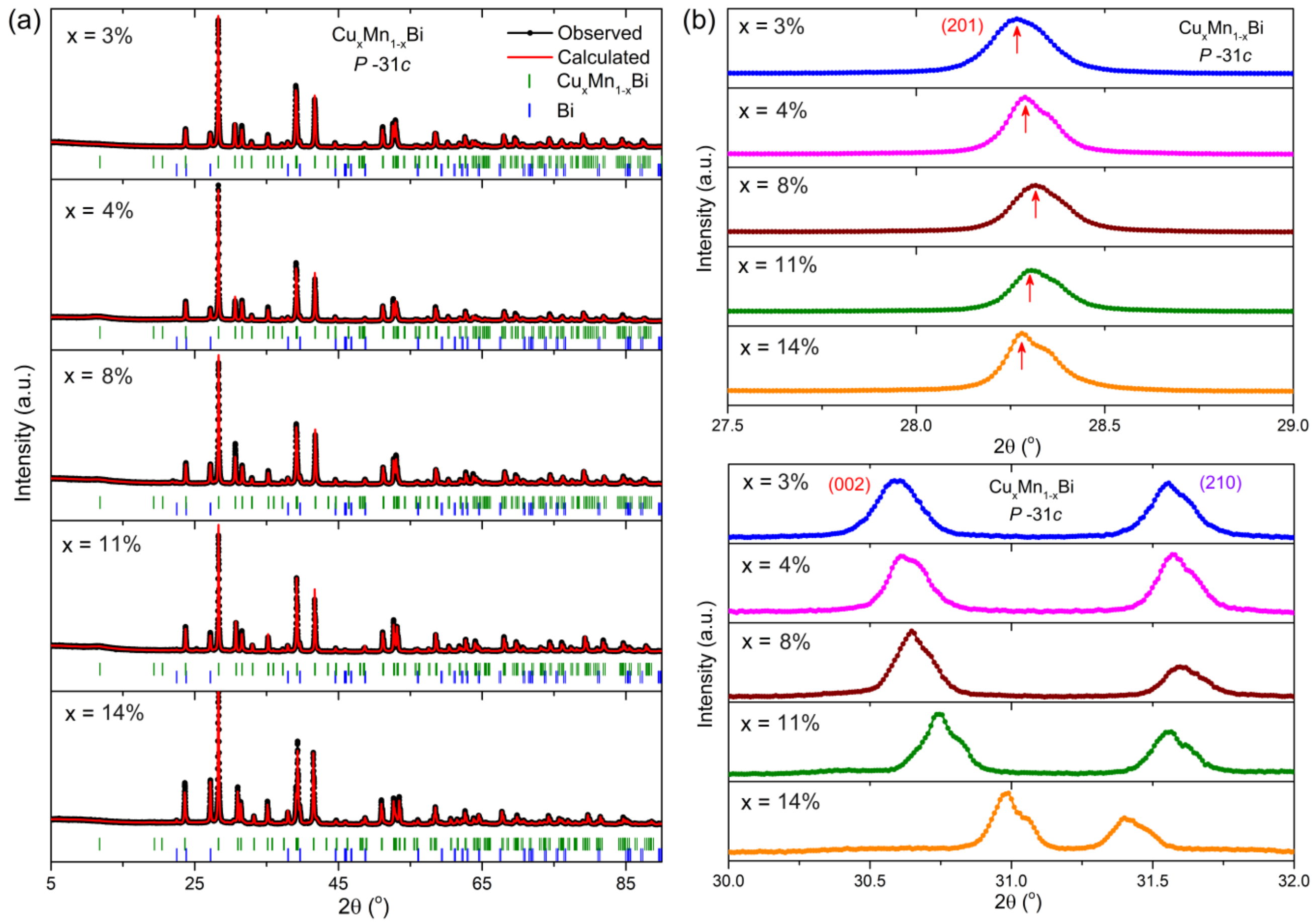

**Figure 3. (a)** Powder X-ray diffraction patterns for $Cu_xMn_{1-x}Bi$. The observed pattern is represented by the black line with dots while the red line stands for calculated pattern. Bragg peaks positions of $Cu_xMn_{1-x}Bi$ and Bi are indicated by green and blue vertical bars, respectively. **(b)** The trends of (201), (002) and (210) peak positions.

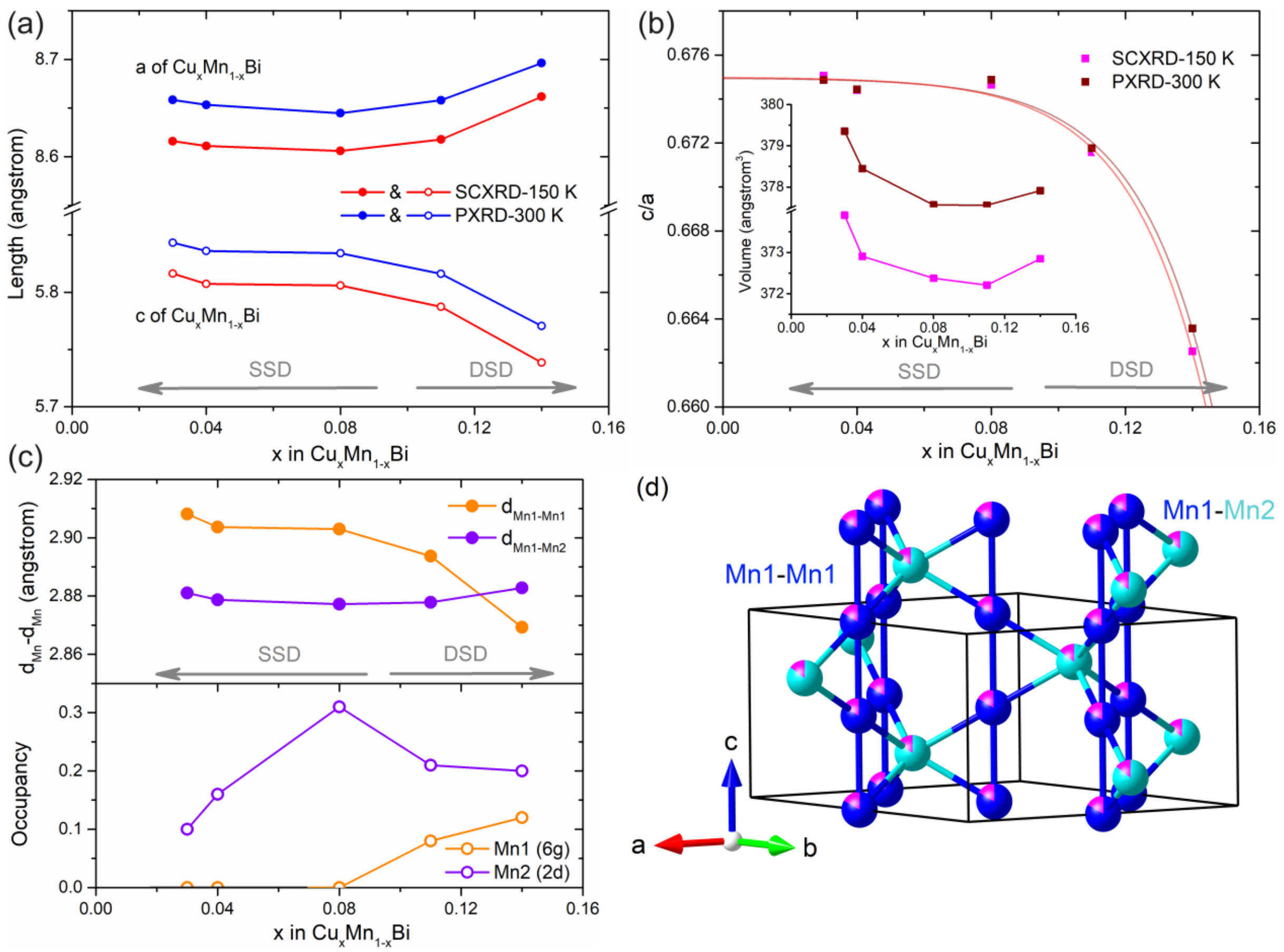

**Figure 4. (a)** The lattice dimensions of $Cu_xMn_{1-x}Bi$ **(b) (Main panel)** *c*/*a* ratio and **(Inset)** unit cell volume **(c)** Mn-Mn interatomic distances and site occupancies of Mn1 and Mn2 sites with respect to x. **(d)** The Cu-doped Mn framework of $Cu_{0.15}Mn_{0.85}Bi$.

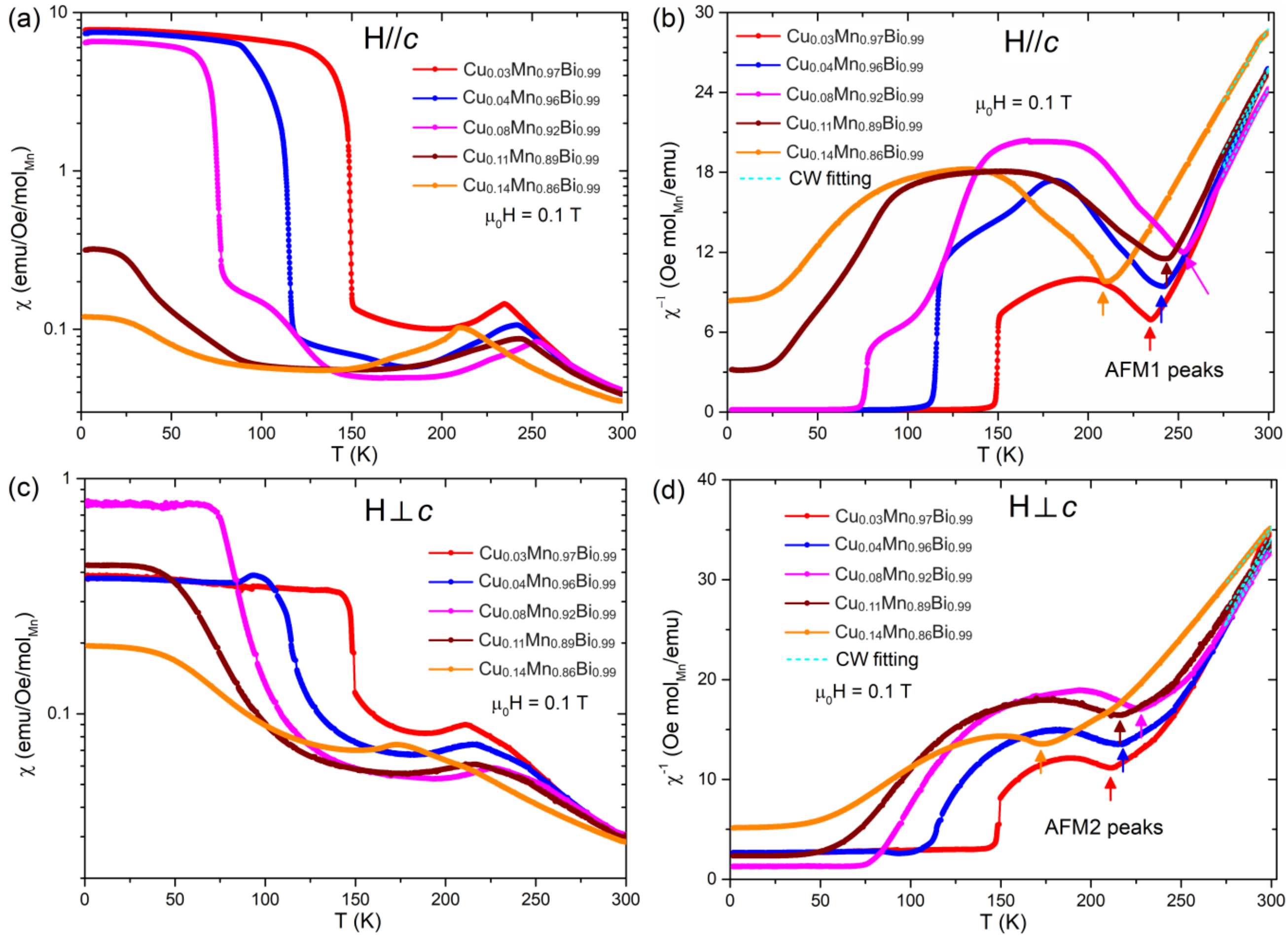

**Figure 5.** Temperature dependence of magnetic susceptibility of $Cu_xMn_{1-x}Bi$ measured under external magnetic field of 0.1 T when **(a)** H//*c* and **(c)** H⊥*c* and the corresponding inverse magnetic susceptibility with respect to temperature when **(b)** H//*c* and **(d)** H⊥*c*. The cyan dashed lines stand for the Curie-Weiss fitting.

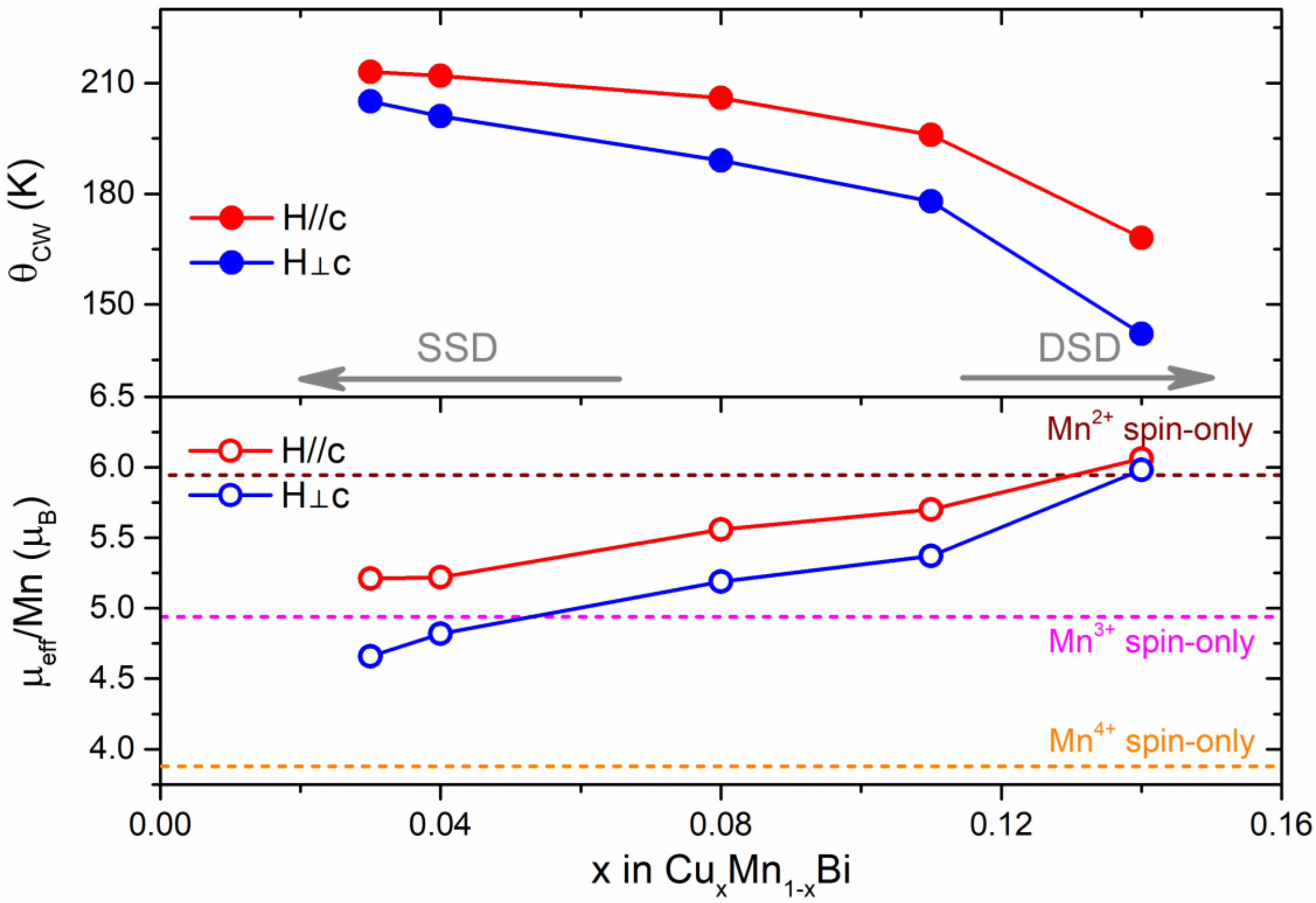

**Figure 6.** Fitted Curie-Weiss temperature ($\theta_{CW}$) and effective moment ($\mu_{eff}$) with respect to x in $Cu_xMn_{1-x}Bi$. The red and blue lines with full/open circles stand for the results when H//*c* and H⊥*c*, respectively.

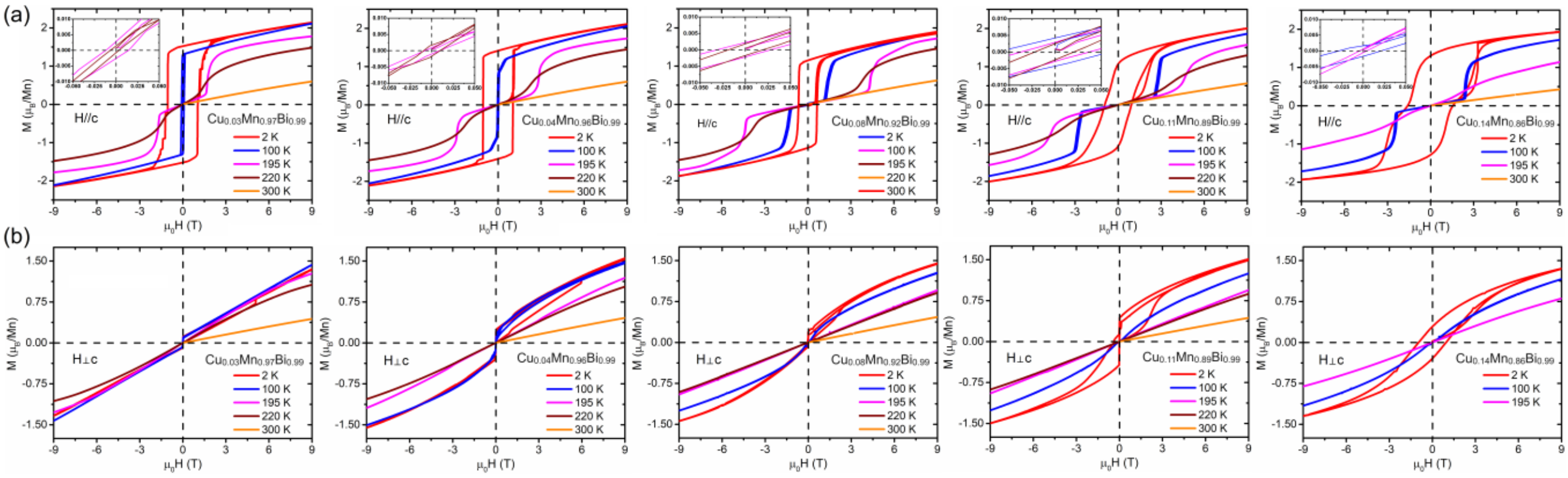

**Figure 7.** Hysteresis loops for $Cu_xMn_{1-x}Bi$ under various temperatures when **(a)** H//*c* and **(b)** H⊥*c*. The **insets** in **(a)** are the enlarged region from -0.05 T to 0.05 T.

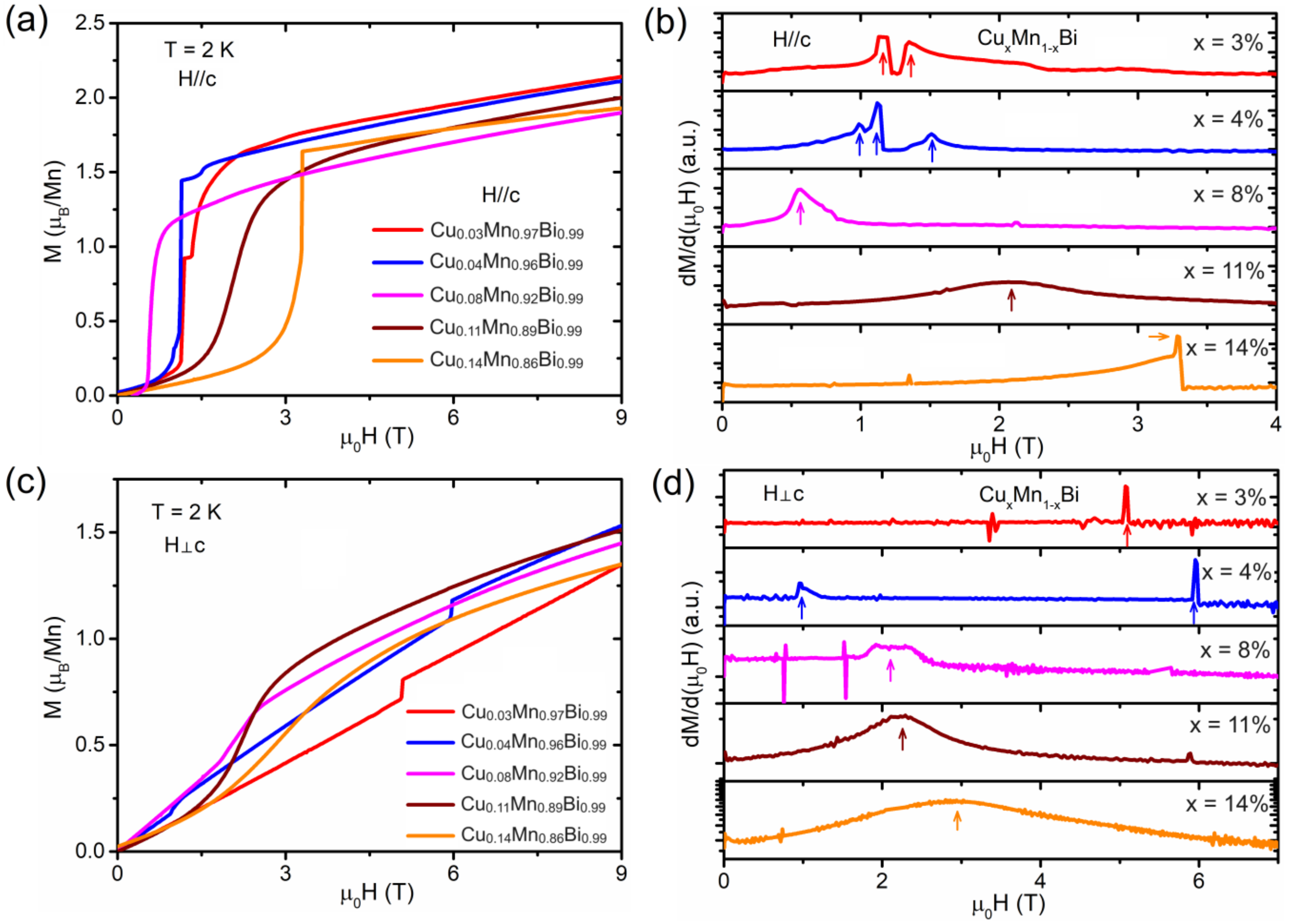

**Figure 8.** Magnetic field dependence of magnetization for $Cu_xMn_{1-x}Bi$ under 2 K when **(a)** H//*c* and **(c)** H⊥*c* and **(b)** & **(d)** the corresponding first derivatives. The arrows in **(b)** & **(d)** indicate the location of the change of slope in the MH curves in **(a)** & **(c)**.

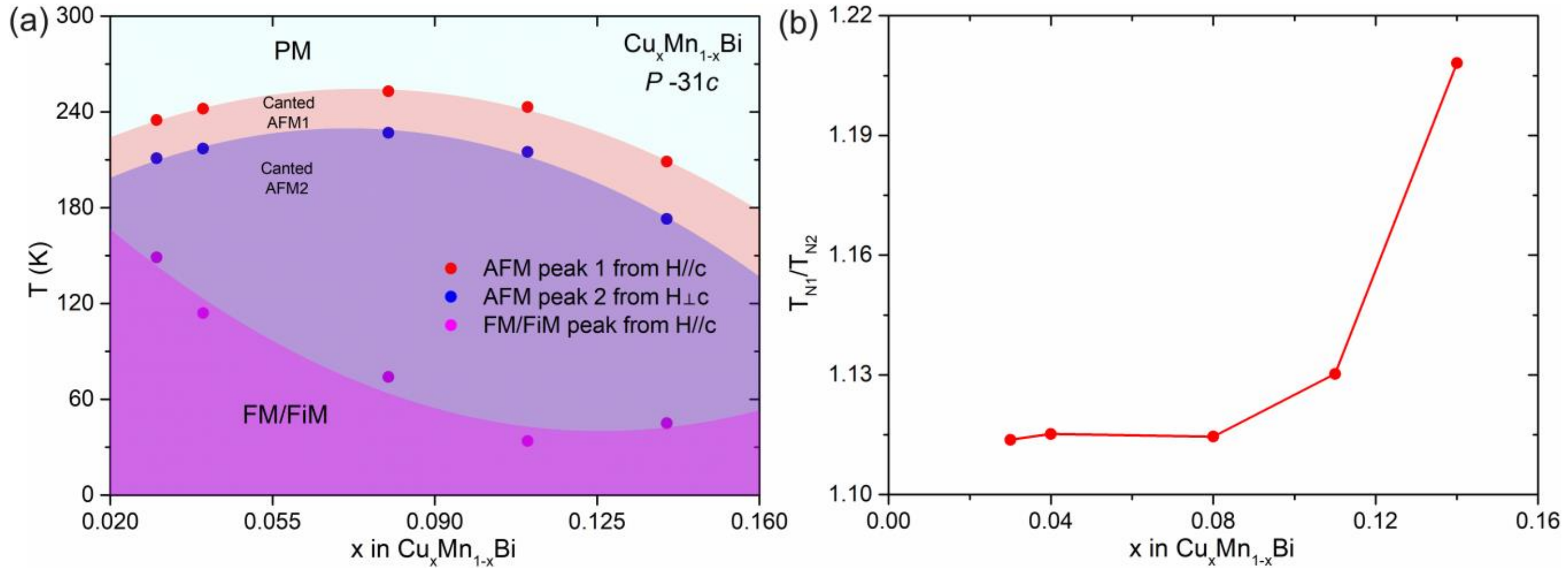

**Figure 9. (a)** Magnetic phase diagram of $Cu_xMn_{1-x}Bi$. Red, blue and purple dots are extracted from AFM peak1, AFM peak2 and FM/FiM peak positions from Figure 5. The boundaries between regions is obtained from a parabola fitting. **(b)** The change of $T_{N1}/T_{N2}$ ratio with respect to x in $Cu_xMn_{1-x}Bi$.

**For Table of Contents Only**

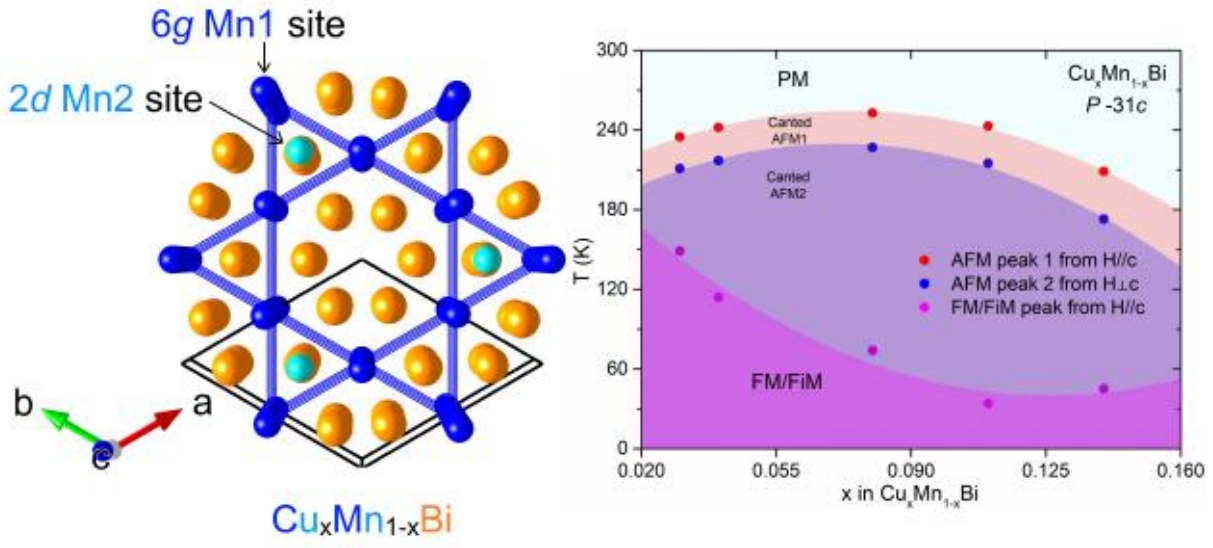

Cu-doped MnBi crystals adopt a new type of high-temperature phase of the well-known permanent magnet, MnBi. The magnetic diagram demonstrates the evolution of magnetic orders along with increasing concentration of Cu in such a system.

# Magnetic Order in A Quenched-High-Temperature-Phase of Cu-Doped MnBi

*Gina Angelo,*[a] *Jeremy G. Philbrick,*[b] *Jian Zhang,*[d] *Tai Kong,*[b,c] *Xin Gui* [a*]

[a] Department of Chemistry, University of Pittsburgh, Pittsburgh, PA, 15260, USA
[b] Department of Physics, University of Arizona, Tucson, AZ, 85721, USA
[c] Department of Chemistry and Biochemistry, University of Arizona, Tucson, AZ, 85721, USA
[d] The Molecular Foundry, Lawrence Berkeley National Laboratory, Berkeley, CA, 94720, USA

## Table of Contents

**Table S1.** Atomic coordinates and equivalent isotropic displacement parameters for $Cu_xMn_{1-x}Bi$ at 150 (2) K. ($U_{eq}$ is defined as one-third of the trace of the orthogonalized $U_{ij}$ tensor (Å$^2$))

$Cu_{0.03(2)}Mn_{0.97(2)}Bi_{0.99(1)}$:

| Atom | Wyck. | Occ. | *x* | *y* | *z* | $U_{eq}$ |
|---|---|---|---|---|---|---|
| Bi1 | 6*h* | 1 | 0.2946 (1) | 0.1473 (1) | ¼ | 0.0129 (1) |
| Bi2 | 2*c* | 0.843 (3) | ⅓ | ⅔ | ¼ | 0.0102 (2) |
| Bi3 | 6*h* | 0.036 (1) | 0.2275 (5) | 0.7725 (5) | ¼ | 0.036 (1) |
| Mn1 | 6*g* | 1 | ½ | 0 | 0 | 0.0134 (3) |
| Mn2 | 2*d* | 0.67 (6) | ⅔ | ⅓ | ¼ | 0.014 (1) |
| Cu1 | 2*d* | 0.10 (6) | ⅔ | ⅓ | ¼ | 0.014 (1) |
| Mn3 | 4*f* | 0.1156 | ⅔ | ⅓ | 0.138 (4) | 0.014 (1) |

$Cu_{0.04(2)}Mn_{0.96(2)}Bi_{0.99(1)}$:

| Atom | Wyck. | Occ. | *x* | *y* | *z* | $U_{eq}$ |
|---|---|---|---|---|---|---|
| Bi1 | 6*h* | 1 | 0.8527 (1) | 0.1473 (1) | ¼ | 0.0129 (2) |
| Bi2 | 2*c* | 0.820 (4) | ⅓ | ⅔ | ¼ | 0.0106 (3) |
| Bi3 | 6*h* | 0.040 (1) | 0.2280 (6) | 0.7720 (6) | ¼ | 0.0106 (5) |
| Mn1 | 6*g* | 1 | ½ | 0 | 0 | 0.0138 (3) |
| Mn2 | 2*d* | 0.49 (7) | ⅔ | ⅓ | ¼ | 0.008 (2) |
| Cu1 | 2*d* | 0.16 (7) | ⅔ | ⅓ | ¼ | 0.008 (2) |
| Mn3 | 4*f* | 0.176 | ⅔ | ⅓ | 0.152 (2) | 0.008 (2) |

$Cu_{0.08(2)}Mn_{0.92(2)}Bi_{0.99(1)}$:

| Atom | Wyck. | Occ. | *x* | *y* | *z* | $U_{eq}$ |
|---|---|---|---|---|---|---|
| Bi1 | 6*h* | 1 | 0.7055 (1) | 0.8527 (1) | ¼ | 0.0140 (2) |
| Bi2 | 2*d* | 0.820 (4) | ⅓ | ⅔ | ¼ | 0.0118 (3) |
| Bi3 | 6*h* | 0.0414 (13) | 0.458 (1) | 0.229 (1) | ¼ | 0.0118 (3) |
| Mn1 | 6*g* | 1 | ½ | 0 | 0 | 0.0143 (3) |
| Mn2 | 2*c* | 0.41 (8) | ⅔ | ⅓ | ¼ | 0.018 (2) |
| Cu1 | 2*c* | 0.31 (8) | ⅔ | ⅓ | ¼ | 0.018 (2) |
| Mn3 | 4*f* | 0.1363 | ⅔ | ⅓ | 0.144 (4) | 0.018 (2) |

$Cu_{0.11(4)}Mn_{0.89(4)}Bi_{0.99(1)}$:

| **Atom** | **Wyck.** | **Occ.** | ***x*** | ***y*** | ***z*** | $U_{eq}$ |
|---|---|---|---|---|---|---|
| Bi1 | 6*h* | 1 | 0.2947 (1) | 0.1473 (1) | ¼ | 0.0149 (2) |
| Bi2 | 2*c* | 0.799 (3) | ⅓ | ⅔ | ¼ | 0.0129 (2) |
| Bi3 | 6*h* | 0.0489 (1) | 0.5409 (7) | 0.7705 (3) | ¼ | 0.0129 (3) |
| Mn1 | 6*g* | 0.92 (3) | ½ | 0 | 0 | 0.0160 (4) |
| Cu1 | 6*g* | 0.08 (3) | ½ | 0 | 0 | 0.0160 (4) |
| Mn2 | 2*d* | 0.43 (5) | ⅔ | ⅓ | ¼ | 0.014 (1) |
| Cu2 | 2*d* | 0.21 (5) | ⅔ | ⅓ | ¼ | 0.014 (1) |
| Mn3 | 4*f* | 0.1808 | ⅔ | ⅓ | 0.148 (2) | 0.014 (1) |

$Cu_{0.14(7)}Mn_{0.86(7)}Bi_{0.99(1)}$:

| **Atom** | **Wyck.** | **Occ.** | ***x*** | ***y*** | ***z*** | $U_{eq}$ |
|---|---|---|---|---|---|---|
| Bi1 | 6*h* | 1 | 0.2960 (1) | 0.1480 (1) | ¼ | 0.0151 (3) |
| Bi2 | 2*c* | 0.738 (5) | ⅓ | ⅔ | ¼ | 0.0143 (4) |
| Bi3 | 6*h* | 0.074 (2) | 0.5389 (9) | 0.7694 (5) | ¼ | 0.0143 (6) |
| Mn1 | 6*g* | 0.88 (6) | ½ | 0 | 0 | 0.0165 (8) |
| Cu1 | 6*g* | 0.12 (6) | ½ | 0 | 0 | 0.0165 (8) |
| Mn2 | 2*d* | 0.25 (9) | ⅔ | ⅓ | ¼ | 0.017 (5) |
| Cu2 | 2*d* | 0.20 (9) | ⅔ | ⅓ | ¼ | 0.017 (5) |
| Mn3 | 4*f* | 0.2754 | ⅔ | ⅓ | 0.154 (3) | 0.017 (5) |

**Table S2.** Anisotropic thermal displacement parameters for $Cu_xMn_{1-x}Bi$ at 150 (2) K.

$Cu_{0.03(2)}Mn_{0.97(2)}Bi_{0.99(1)}$:

| Atom | $U_{11}$ | $U_{22}$ | $U_{33}$ | $U_{23}$ | $U_{13}$ | $U_{12}$ |
|---|---|---|---|---|---|---|
| Bi1 | 0.0101 (2) | 0.0128 (2) | 0.0148 (2) | -0.0000 (1) | 0 | 0.0050 (1) |
| Bi2 | 0.0091 (2) | 0.0091 (2) | 0.0123 (3) | 0 | 0 | 0.0046 (1) |
| Bi3 | 0.0091 (5) | 0.0091 (5) | 0.0123 (2) | 0 | 0 | 0.0046 (3) |
| Mn1 | 0.0148 (6) | 0.0167 (6) | 0.0094 (5) | 0.0021 (4) | 0.0006 (4) | 0.0083 (5) |
| Mn2 | 0.0095 (8) | 0.0095 (8) | 0.024 (2) | 0 | 0 | 0.0048 (4) |
| Cu1 | 0.0095 (8) | 0.0095 (8) | 0.024 (2) | 0 | 0 | 0.0048 (4) |
| Mn3 | 0.0095 (8) | 0.0095 (8) | 0.024 (2) | 0 | 0 | 0.0048 (4) |

$Cu_{0.04(2)}Mn_{0.96(2)}Bi_{0.99(1)}$:

| Atom | $U_{11}$ | $U_{22}$ | $U_{33}$ | $U_{23}$ | $U_{13}$ | $U_{12}$ |
|---|---|---|---|---|---|---|
| Bi1 | 0.0118 (2) | 0.0118 (2) | 0.0173 (2) | 0.0000 (1) | 0.0000 (1) | 0.0075 (2) |
| Bi2 | 0.0076 (3) | 0.0076 (3) | 0.0165 (4) | 0 | 0 | 0.0038 (2) |
| Bi3 | 0.0076 (6) | 0.0076 (6) | 0.0165 (3) | 0 | 0 | 0.0038 (3) |
| Mn1 | 0.0135 (7) | 0.0161 (7) | 0.0127 (7) | 0.0019 (5) | 0.0012 (5) | 0.0081 (6) |
| Mn2 | 0.009 (1) | 0.009 (1) | 0.008 (3) | 0 | 0 | 0.0042 (5) |
| Cu1 | 0.009 (1) | 0.009 (1) | 0.008 (3) | 0 | 0 | 0.0042 (5) |
| Mn3 | 0.009 (1) | 0.009 (1) | 0.008 (3) | 0 | 0 | 0.0042 (5) |

$Cu_{0.08(2)}Mn_{0.92(2)}Bi_{0.99(1)}$:

| Atom | $U_{11}$ | $U_{22}$ | $U_{33}$ | $U_{23}$ | $U_{13}$ | $U_{12}$ |
|---|---|---|---|---|---|---|
| Bi1 | 0.0100 (2) | 0.0134 (2) | 0.0175 (3) | 0.0001 (1) | 0 | 0.0050 (1) |
| Bi2 | 0.0091 (3) | 0.0091 (3) | 0.0172 (4) | 0 | 0 | 0.0045 (2) |
| Bi3 | 0.0091 (7) | 0.0091 (7) | 0.0172 (3) | 0 | 0 | 0.0045 (3) |
| Mn1 | 0.0140 (7) | 0.0170 (8) | 0.0130 (8) | -0.0022 (5) | -0.0009 (5) | 0.0087 (6) |
| Mn2 | 0.010 (1) | 0.010 (1) | 0.034 (3) | 0 | 0 | 0.0051 (6) |
| Cu1 | 0.010 (1) | 0.010 (1) | 0.034 (3) | 0 | 0 | 0.0051 (6) |
| Mn3 | 0.010 (1) | 0.010 (1) | 0.034 (3) | 0 | 0 | 0.0051 (6) |

$Cu_{0.11(4)}Mn_{0.89(4)}Bi_{0.99(1)}$:

| **Atom** | **$U_{11}$** | **$U_{22}$** | **$U_{33}$** | **$U_{23}$** | **$U_{13}$** | **$U_{12}$** |
|---|---|---|---|---|---|---|
| Bi1 | 0.0109 (2) | 0.0148 (1) | 0.0176 (2) | 0.0000 (1) | 0 | 0.0055 (1) |
| Bi2 | 0.0101 (2) | 0.0101 (2) | 0.0185 (3) | 0 | 0 | 0.0051 (1) |
| Bi3 | 0.0101 (4) | 0.0101 (4) | 0.0185 (2) | 0 | 0 | 0.0051 (2) |
| Mn1 | 0.0164 (6) | 0.0193 (6) | 0.0137 (6) | 0.0023 (4) | 0.0013 (4) | 0.0099 (4) |
| Cu1 | 0.0164 (6) | 0.0193 (6) | 0.0137 (6) | 0.0023 (4) | 0.0013 (4) | 0.0099 (4) |
| Mn2 | 0.0103 (7) | 0.0103 (7) | 0.020 (2) | 0 | 0 | 0.0052 (4) |
| Cu2 | 0.0103 (7) | 0.0103 (7) | 0.020 (2) | 0 | 0 | 0.0052 (4) |
| Mn3 | 0.0103 (7) | 0.0103 (7) | 0.020 (2) | 0 | 0 | 0.0052 (4) |

$Cu_{0.14(7)}Mn_{0.86(7)}Bi_{0.99(1)}$:

| **Atom** | **$U_{11}$** | **$U_{22}$** | **$U_{33}$** | **$U_{23}$** | **$U_{13}$** | **$U_{12}$** |
|---|---|---|---|---|---|---|
| Bi1 | 0.0088 (4) | 0.0135 (3) | 0.0214 (4) | -0.0001 (2) | 0 | 0.0044 (2) |
| Bi2 | 0.0090 (5) | 0.0090 (5) | 0.0251 (7) | 0 | 0 | 0.0045 (3) |
| Bi3 | 0.0090 (5) | 0.0090 (5) | 0.0251 (7) | 0 | 0 | 0.0045 (3) |
| Mn1 | 0.016 (1) | 0.020 (1) | 0.016 (1) | 0.004 (1) | 0.002 (1) | 0.011 (1) |
| Cu1 | 0.016 (1) | 0.020 (1) | 0.016 (1) | 0.004 (1) | 0.002 (1) | 0.011 (1) |
| Mn2 | 0.009 (2) | 0.009 (2) | 0.033 (16) | 0 | 0 | 0.003 (1) |
| Cu2 | 0.009 (2) | 0.009 (2) | 0.033 (16) | 0 | 0 | 0.003 (1) |
| Mn3 | 0.009 (2) | 0.009 (2) | 0.033 (16) | 0 | 0 | 0.003 (1) |

**Figure S1.** Single crystal X-ray diffraction patterns from (0kl), (h0l) and (hk0) planes for $Cu_{0.03(2)}Mn_{0.97(2)}Bi_{0.99(1)}$.

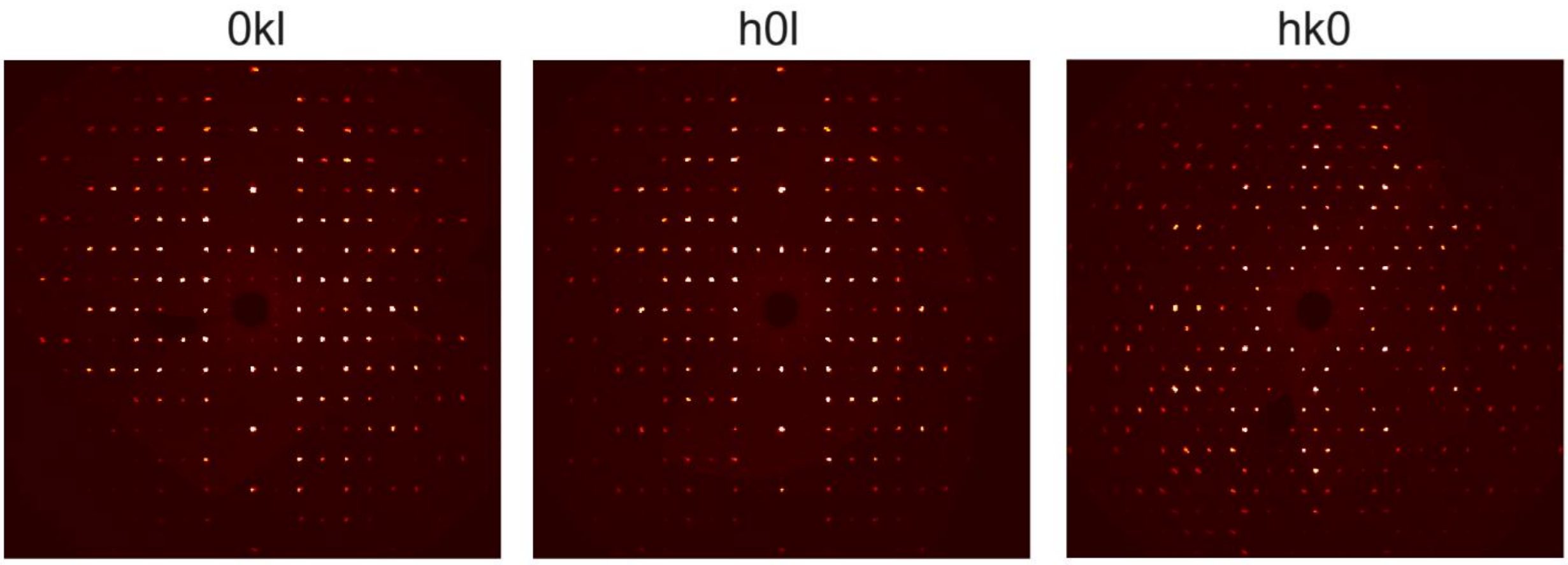

**Figure S2.** Single crystal X-ray diffraction patterns from (0kl), (h0l) and (hk0) planes for $Cu_{0.04(2)}Mn_{0.96(2)}Bi_{0.99(1)}$.

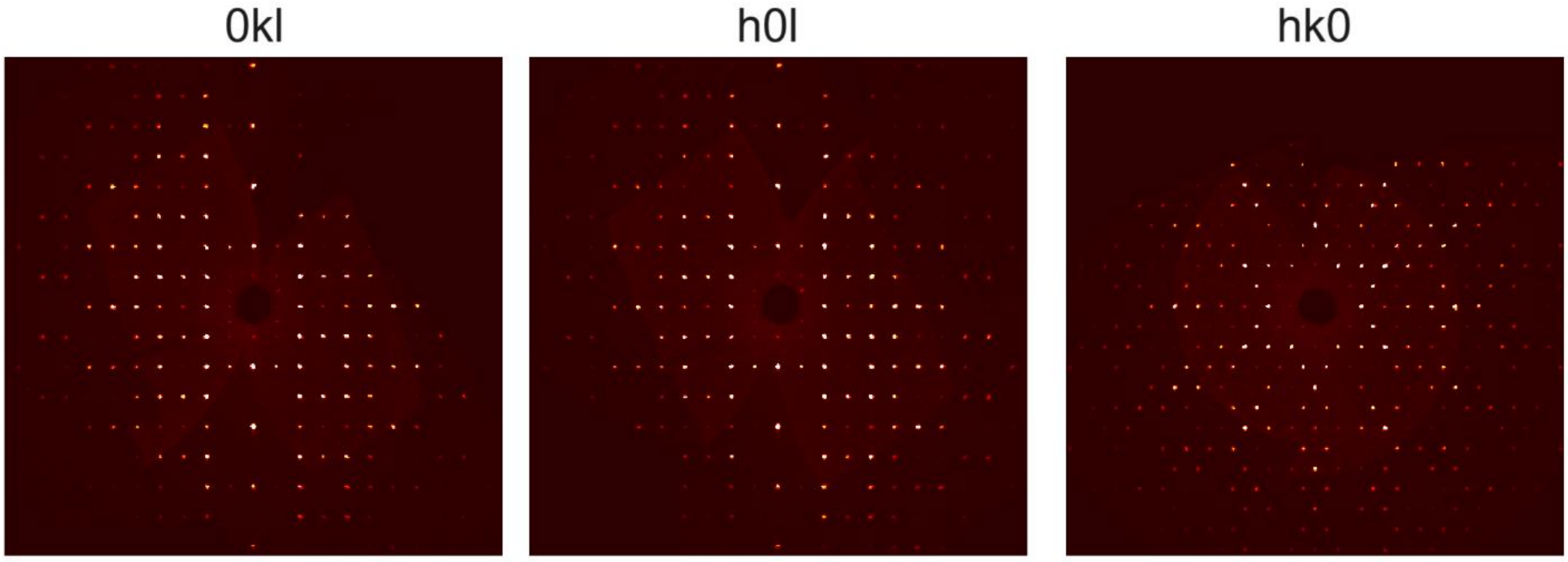

**Figure S3.** Single crystal X-ray diffraction patterns from (0kl), (h0l) and (hk0) planes for $Cu_{0.08(2)}Mn_{0.92(2)}Bi_{0.99(1)}$.

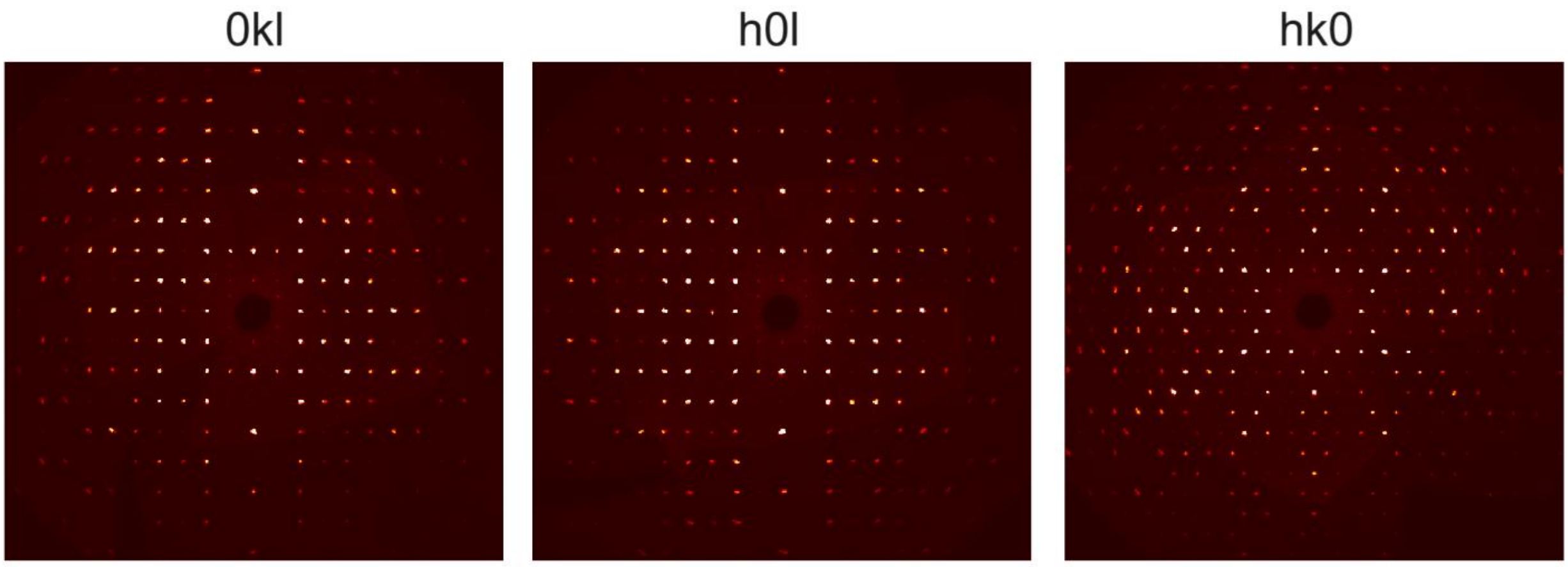

**Figure S4.** Single crystal X-ray diffraction patterns from (0kl), (h0l) and (hk0) planes for $Cu_{0.11(4)}Mn_{0.89(4)}Bi_{0.99(1)}$.

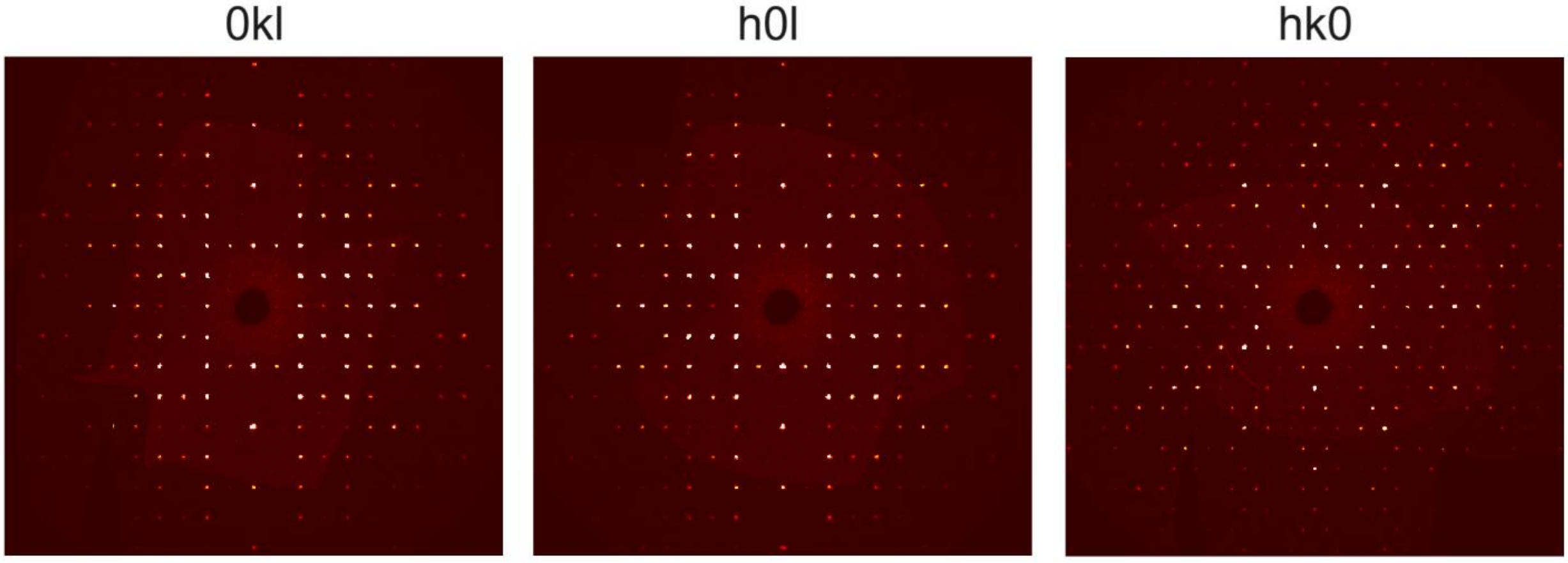

**Figure S5.** Single crystal X-ray diffraction patterns from (0kl), (h0l) and (hk0) planes for $Cu_{0.14(7)}Mn_{0.86(7)}Bi_{0.99(1)}$.

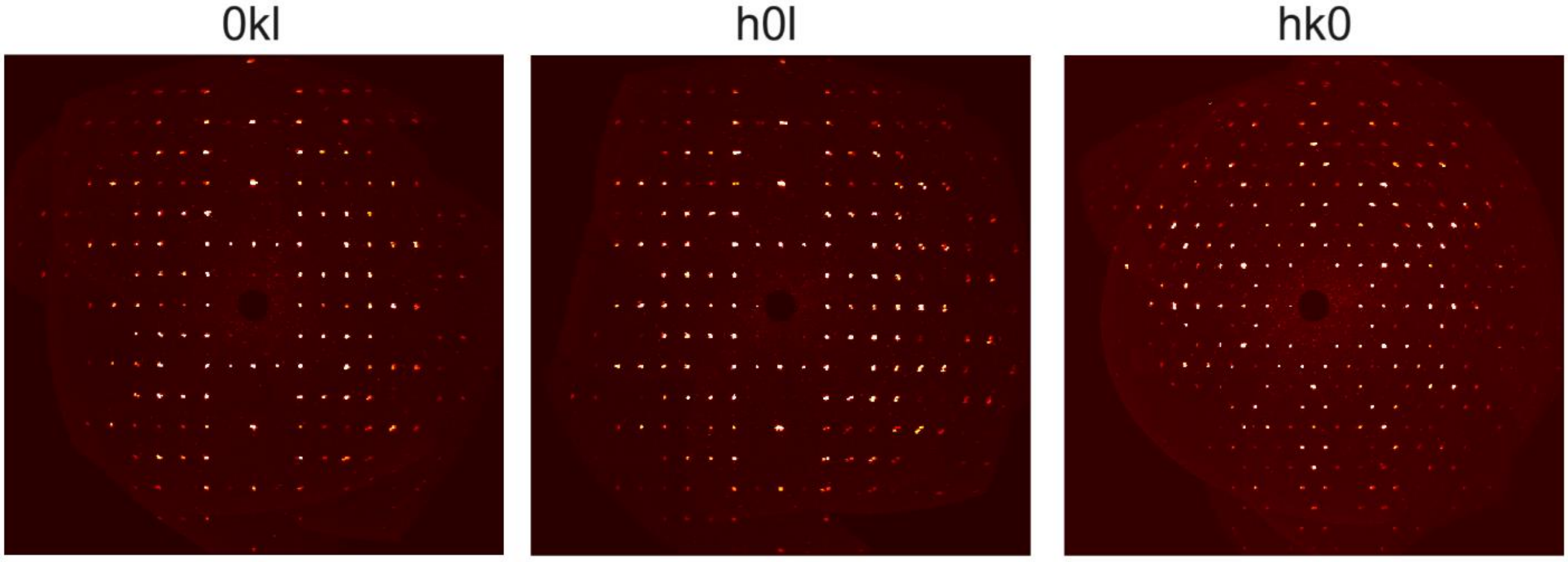

**Figure S6.** Comparison of powder XRD patterns for $Cu_xMn_{1-x}Bi$ and reported QHTP-MnBi.

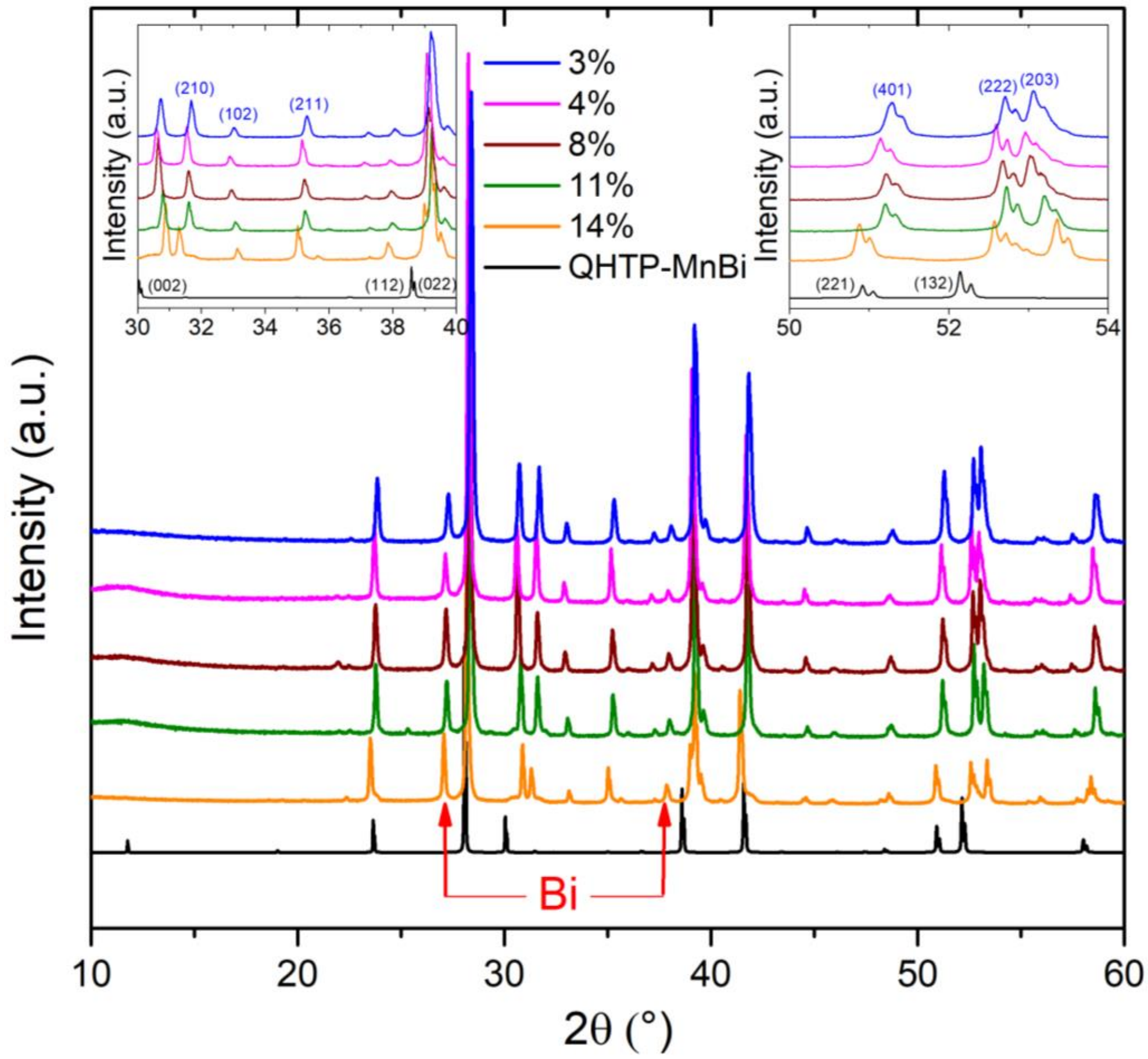

**Figure S7.** Temperature-dependent heat capacity ($C_p$) of $Cu_{0.04}Mn_{0.96}Bi_{0.99}$.

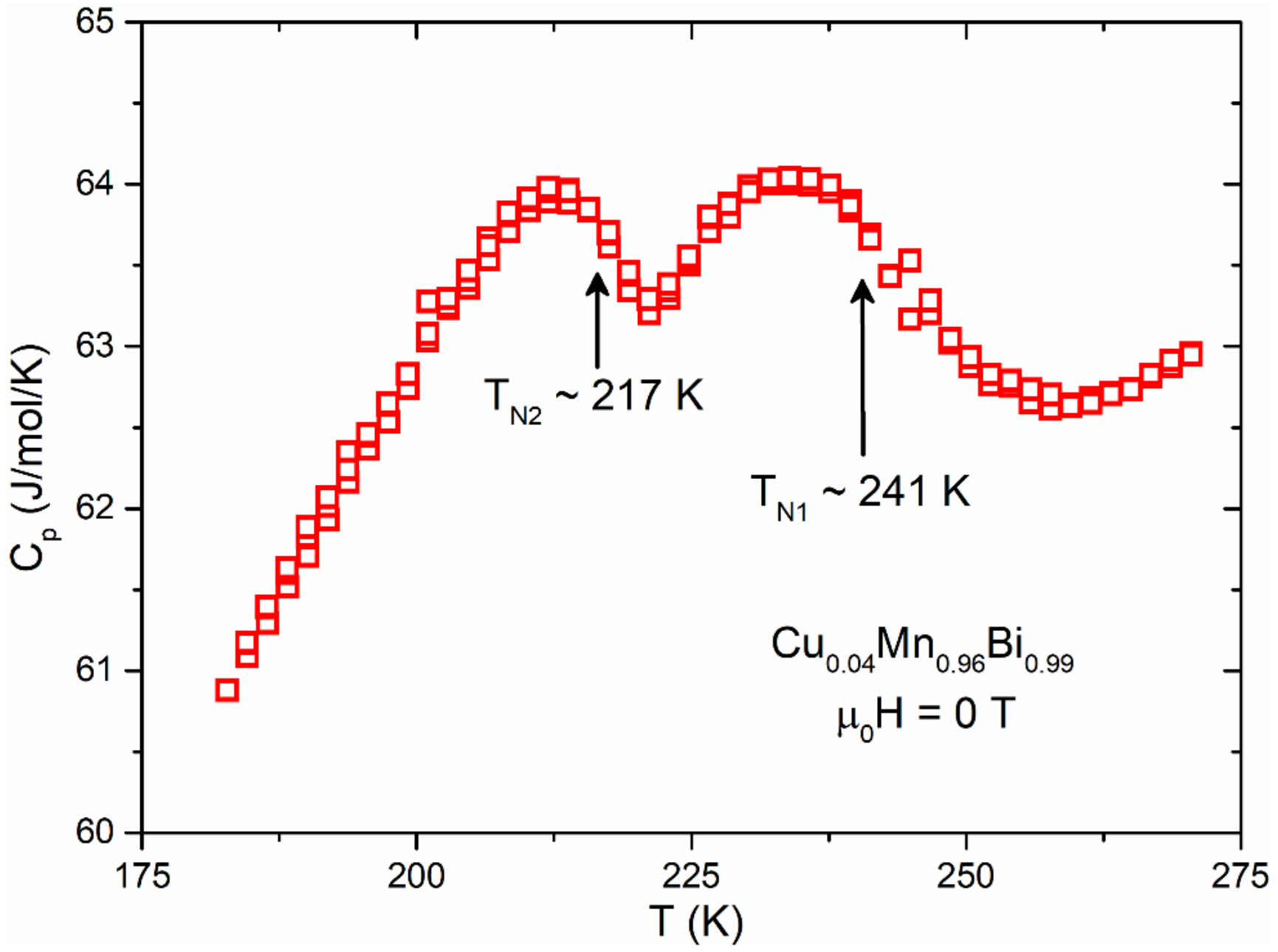

**Figure S8.** The first derivative of $\chi T$ vs T curves for **(a)** $Cu_{0.11(4)}Mn_{0.89(4)}Bi_{0.99(1)}$ and **(b)** $Cu_{0.14(7)}Mn_{0.86(7)}Bi_{0.99(1)}$.

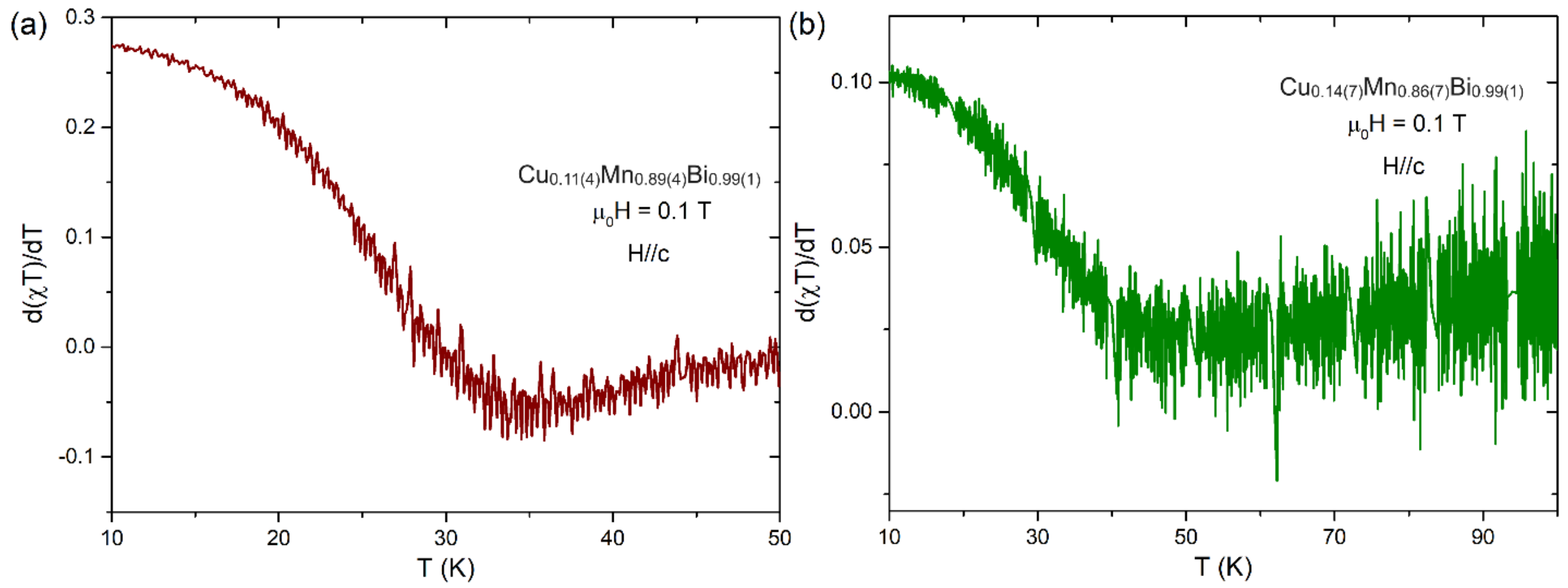